\documentclass[preprint,12pt]{elsarticle}

\usepackage{amssymb}
\usepackage{amsmath}
\usepackage{hyperref}
\usepackage[ruled,vlined]{algorithm2e}
\RestyleAlgo{ruled}
\usepackage{float}

\usepackage{booktabs}

\usepackage{graphicx}
\usepackage{subfigure}

\usepackage{tcolorbox}
\usepackage[section]{placeins}
\usepackage{tikz}
\usetikzlibrary{arrows.meta,positioning,shapes,fit}
\usepackage{listings}
\usepackage{xcolor}

\journal{Applied Energy}

\begin{document}

\begin{frontmatter}



\title{LLM-Guided Safe Reinforcement Learning for Energy System Topology Reconfiguration}


\author[label1]{Zongyan Zhang}
\author[label2]{Chao Shen}
\author[label2]{Xu Wan}
\author[label1]{Jie Song}
\author[label1]{Mingyang Sun \corref{cor1}} 
\cortext[cor1]{Corresponding Author}
\affiliation[label1]{organization={School of Advanced Manufacturing and Robotics,Peking University},
            postcode={100871},
            city={Beijing},
            country={China}}

\affiliation[label2]{organization={Collage of Control Science and Engineering, Zhejiang University},
            postcode={310027},
            city={Hangzhou},
            country={China}}

\begin{abstract}
The increasing penetration of renewable generation and the growing variability of electrified demand introduce substantial operational uncertainty to modern power systems. Topology reconfiguration is widely recognized as an effective and economical means to enhance grid resilience. Due to the coexistence of AC power-flow constraints and discrete switching decisions, topology reconfiguration in large‐scale systems leads to a highly nonlinear and nonconvex optimization problem, making traditional methods computationally prohibitive. Consequently, several studies have explored reinforcement learning–based approaches to improve scalability and operational efficiency. However, its practical implementation is challenged by the high-dimensional combinatorial action space and the need to ensure safety during learning-based decision-making. To address these challenges, this paper presents a safe and intelligent topology control framework that integrates Large Language Models (LLMs) with a Safety Soft Actor--Critic (\emph{Safety-SAC}) architecture. Operational voltage and thermal limits are reformulated into smooth safety-cost signals, enabling risk-aware policy optimization within a constrained Markov decision process. A knowledge-based \emph{Safety-LLM} module is further introduced to refine unsafe or suboptimal transitions through domain knowledge and state-informed reasoning, thus guiding the learning agent toward safer and more effective switching actions. Experiments on the IEEE 36-bus and 118-bus Grid2Op benchmarks show that the proposed method consistently improves reward, survival time, and safety metrics, achieving higher reward, longer survival, and lower safety cost compared with SAC, ACE, and their safety-enhanced variants. These results demonstrate the potential of combining LLM-based reasoning with
safe reinforcement learning to achieve scalable and reliable grid topology
control. \textcolor{blue}{In addition, representative examples illustrate how
LLM-generated refinements can be examined at the transition level, thereby
enabling a clearer inspection of corrective control actions without altering
the underlying environment dynamics.}
\end{abstract}




\end{frontmatter}





\section{Introduction}

With the global transition toward deep decarbonization, power systems are undergoing rapid growth in renewable generation and the electrification of transportation and heating sectors \cite{cui2023online, shen2025Physics-Augmented}. While these transitions are essential for achieving long-term emission targets, they introduce substantial operational challenges \cite{shen2025physics}. The intermittency and limited dispatchability of renewable generation, coupled with increasingly volatile electrified loads, significantly increase operational uncertainty. These factors further intensify net-load fluctuations and contribute to higher system peak demand \cite{zeng2022resilience, wan2023adapsafe}. Under such evolving conditions, ensuring secure and reliable grid operation has become increasingly difficult, motivating the development of advanced data-driven and intelligence-augmented operational frameworks \cite{oskouei2022resilience}.

Grid operators increasingly view dynamic topology reconfiguration as a cost-effective and flexible complement to traditional operational measures, such as generation curtailment, load shedding, peak shaving, and transmission expansion planning \cite{fisher2008optimal, peker2018benefits, si2024multi}. Although these reconfiguration actions are technically feasible, achieving optimal topology control remains challenging, particularly in large-scale systems \cite{matavalam2022curriculum}. In current practice, operational decisions primarily focus on transmission line switching, while more advanced actions, including bus splitting or bus merging within substations, involve a far larger set of feasible topological configurations \cite{si2024multi}. The resulting combinatorial growth of the decision space, coupled with the nonlinear structure of power network physics, renders large-scale topology reconfiguration computationally prohibitive under real-time operational constraints \cite{lai2022network, oh2020online}.

In \cite{khanabadi2012optimal}, transmission line switching is embedded into the AC optimal power flow (ACOPF) problem by introducing binary decision variables, resulting in a mixed integer nonlinear programming (MINLP) formulation solved through Benders decomposition. While this approach captures both economic efficiency and operational security, it does not consider the long-term effects of switching actions over a planning horizon. Subsequent studies therefore focused on dynamic topology reconfiguration, where optimal switching sequences are determined across multiple time periods.
For example, \cite{granelli2006optimal} introduces a temporal optimization horizon but relies on deterministic and evolutionary MINLP solvers, leading to significant computational overhead. Likewise, \cite{schnyder2002security} examines the identification of an optimal and $N-1$ secure topology and proposes a linearized topology-optimizing power flow method that integrates contingency constraints and corrective switching decisions within a unified framework. Despite these advancements, the resulting optimization models are inherently nonlinear and nonconvex due to the AC power flow constraints and the inclusion of discrete switching decisions, which collectively lead to severe computational complexity in large-scale settings \cite{cui2023online}.

Recent research has increasingly explored deep reinforcement learning (DRL) as a means to enable intelligent topology control in power systems. Reinforcement learning (RL) provides a data-driven framework in which an agent improves its decision-making policy through repeated interaction with the environment and performance-based feedback. In \cite{cui2023online}, the preventive control problem for transmission-overload (TO) relief is formulated as a constrained Markov decision process (CMDP) that jointly coordinates generation redispatch, transmission switching, and busbar switching. The resulting CMDP is solved using the interior-point policy optimization (IPO) algorithm, ensuring safe exploration under operational constraints. {\color{red}Although CMDP-based formulations provide a principled way to incorporate operational constraints, their practical implementation in topology reconfiguration remains challenging.} The work in \cite{si2024multi} develops a multi-agent reinforcement learning architecture with an attention-based dynamic agent network and an action-masking mechanism to handle sequential distribution system restoration under dynamic network reconfiguration. In \cite{GaoCloud-Edge}, a cloud-edge collaborative framework is proposed for large-scale distribution network reconfiguration, employing a discrete multi-agent soft actor-critic (MASAC) approach together with safe offline and online learning mechanisms to enable fast, scalable, and constraint-compliant switching decisions. In \cite{subramanian2021exploring}, a DRL framework based on the Cross-Entropy Method (CEM) is used to train an artificial control-room agent for topology reconfiguration to relieve line overloads and maintain secure power flows under varying grid conditions. Despite these advances, dynamic topology reconfiguration remains challenging, as the state and action spaces grow exponentially with system size due to the combinatorial structure of power networks. This high dimensionality hampers learning efficiency, slows convergence, and often leads to suboptimal policies, particularly when physical system knowledge is not incorporated to guide exploration \cite{matavalam2022curriculum, dwivedi2024blackout, yoon2021winning}.

To improve learning efficiency in high-dimensional decision spaces, recent work has explored physics-guided reinforcement learning (RL) methods that incorporate power-system priors or structured training curricula. In \cite{matavalam2022curriculum}, a curriculum-based asynchronous advantage actor-critic (A3C) framework is proposed for topology reconfiguration, integrating physics-guided action pruning, state abstraction via power-flow correlations, and line-current-based reward shaping to enable faster and more stable convergence. The study in \cite{dwivedi2024blackout} develops a physics-informed deep Q-learning approach for blackout mitigation, where power transfer distribution factors (PTDFs) and line outage distribution factors (LODFs) guide the agent toward physically consistent and security-aware actions. In \cite{yoon2021winning}, a Semi-Markov Afterstate Actor-Critic (SMAAC) architecture is introduced, combining afterstate representations, goal-conditioned hierarchical policies, and graph neural networks to compactly encode equivalent topological configurations and mitigate the combinatorial growth of state-action spaces. Despite these advances, existing physics-guided RL approaches still face two key limitations. First, the physical priors used in these methods are static and manually specified, typically based on fixed sensitivities \cite{dwivedi2024blackout} or predefined pruning rules \cite{matavalam2022curriculum}, and therefore cannot adapt to changing operating conditions or capture the full nonlinear behavior of AC power flows. Second, these approaches do not explicitly enforce operational safety during exploration, leaving intermediate learning trajectories potentially outside secure operating limits and limiting their suitability for real-time use.

Considering the strong knowledge and reasoning capabilities demonstrated by large language models (LLMs), recent studies have explored using LLMs as expert surrogates to guide reinforcement learning (RL) in power system applications \cite{wan2025think, zhang2025domain, feng2025implementing}. Their findings indicate that LLMs can acquire extensive domain knowledge and reasoning patterns from large-scale technical corpora \cite{wan2025think, zhang2025domain, feng2025implementing}. Building on this foundation, LLMs can draw on engineering literature, operational guidelines, and technical documentation to form a broad contextual understanding of power system principles, fault behaviors, and operational constraints \cite{deng2025power, yang2025large}. Such pretrained knowledge enables LLMs to provide adaptive, context-aware guidance for RL exploration without requiring explicit analytical models \cite{liu2024rl}. Moreover, their semantic reasoning capabilities support the interpretation of complex system conditions and facilitate the generation of knowledge-based control intent that links operational objectives to actionable switching decisions \cite{bai2023qwen, guo2025deepseek}. \textcolor{blue}{Integrating LLMs into RL therefore offers a promising pathway to overcome the limitations of static heuristics while enhancing exploration efficiency and providing greater transparency in decision refinement for dynamic grid reconfiguration tasks.}

To ensure safer collaboration between LLMs and RL in topology adjustment decision-making, we propose an LLM-enhanced safe reinforcement learning framework for dynamic topology reconfiguration in power systems. This framework is designed to address the challenges arising from large combinatorial action spaces and the stringent requirements of safe exploration.

The main contributions of this work are summarized as follows:

\begin{enumerate}
    \item To the best of the authors' knowledge, this is the first study that integrates LLM with a safe RL paradigm for power system topology reconfiguration. A unified framework is established that combines safety-constrained policy learning with LLM-based reasoning, enabling adaptive, interpretable, and safety-aware decision-making in large-scale topology control problems.
    \item To tackle the challenge posed by large combinatorial action spaces under strict operational limits, we develop a safety-constrained reinforcement learning formulation tailored to topology reconfiguration. Hard voltage and thermal limits are transformed into differentiable safety-cost signals, forming a bi-objective learning structure optimized through a dual-critic Safety Soft Actor-Critic framework, ensuring consistent reward and safety-value estimation.
    \item Through structured prompts, the LLM analyzes the system state, identifies potential issues, and then suggests physics-informed switching adjustments. These refined transitions are inserted into the replay buffer to guide exploration toward safer and more effective regions of the action space.
    \item Experiments on IEEE 36-bus and 118-bus systems show that the proposed method consistently reduces overloads and voltage violations, extends survival time, and increases reward beyond SAC, ACE, and safety-enhanced baselines, demonstrating superior safety and performance.
\end{enumerate}

\section{Preliminary}
\subsection{Problem Formulation for Optimal Topology Control}
\label{sec:2.1}
The objective of this study is to determine an optimal grid topology over the time horizon $t = \{1, 2, \dots, n\}$ that minimizes the ratio of total load to generation, while also considering the cost associated with adjusting the generation of power plants and maximizing the overall system survivability. The optimization problem is formulated as follows:
\begin{equation}
\min_{\tau} \sum_{t=1}^{n} \sum_{p \in \mathcal{E}} \left( \frac{L_{p,t}}{P_{p,t}} \right) + \sum_{t=t_{\text{over}}}^{t_{\text{end}}} \text{penalty} \cdot \Delta t \cdot (P_{p,t} - P_{p,t}^{\text{ref}})^2,
\label{eq:obj}
\end{equation}
subject to
\begin{align}
&f_{\tau}(x_t) = 0, && \forall t \in \{1,2,\dots,n\}, 
\label{eq:pf_constraint} \\[2mm]
&T(x_t) \in A(T(x_{t-1})), && \forall t \in \{2,\dots,n\},
\label{eq:topo_transition} \\[2mm]
&I_{p,t}\le I_{p,t,\max}, && \forall p \in \mathcal{E},~t \in \{1,\dots,n\},
\label{eq:current_limit} \\[2mm]
&V_{p,t}\in [V_{p,\min},V_{p,\max}], && \forall p \in \mathcal{E},~t \in \{1,\dots,n\}.
\label{eq:voltage_limit}
\end{align}

In \eqref{eq:obj}, the first term minimizes the cumulative ratio between load and generation across all buses and time steps, thereby reducing network losses and improving long-term stability. The second term represents the survivability penalty, where $ \text{penalty} $ is a constant factor that penalizes the system for shorter operational lifetimes. The equality constraint \eqref{eq:pf_constraint} enforces the AC power-flow balance under topology configuration $ \tau $ and state vector $ x_t $. The constraint \eqref{eq:topo_transition} restricts feasible topology transitions between consecutive time steps, ensuring operational continuity. The inequality \eqref{eq:current_limit} imposes the thermal limit on each transmission line to prevent overloads, while \eqref{eq:voltage_limit} maintains bus voltages within allowable bounds to guarantee secure and stable grid operation.

\subsection{Soft Actor-Critic (SAC) Algorithm}
\label{sec:2.2}
The Soft Actor-Critic (SAC) algorithm is an off-policy reinforcement learning method based on the maximum entropy framework. The objective of SAC is to optimize the policy by maximizing both the expected return and the entropy of the policy. This maximization encourages exploration by keeping the policy stochastic while also seeking high rewards. The overall objective function of SAC is formulated as:

\begin{equation}
J(\pi) = \mathbb{E}_{s_t, a_t \sim \pi}\left[ Q_{\theta}(s_t, a_t) - \alpha \log \pi_{\theta}(a_t | s_t) \right]
\end{equation}

where $ Q_{\theta}(s_t, a_t) $ represents the Q-value function for the current state-action pair, and $ \pi_{\theta}(a_t | s_t) $ is the policy that selects actions based on the current state $ s_t $. The parameter $ \alpha $ controls the trade-off between exploration and exploitation by adjusting the weight of the entropy term. The goal is to learn a policy that not only maximizes long-term rewards but also explores the environment effectively by maintaining a certain level of randomness in action selection. {\color{red}Intuitively, the maximum-entropy formulation prevents premature convergence to deterministic policies by explicitly encouraging stochastic exploration. This property is particularly beneficial in large combinatorial action spaces, where insufficient exploration may trap the agent in suboptimal switching patterns.}

\subsubsection{Value Function Loss}

To address the issue of overestimation bias in Q-values, SAC employs Double Q-learning, which involves two Q-value networks. The value function loss function $ L_Q $ is defined as:

\begin{equation}
L_Q(\phi) = \mathbb{E}_{s, a, r, s'} \left[ \left( Q_{\phi}(s, a) - \left( r + \gamma \min_{i=1,2} Q_{\phi_i}(s', a') - \alpha \log \pi_{\theta}(a'|s') \right) \right)^2 \right]
\end{equation}

In this equation, $ Q_{\phi}(s, a) $ is the Q-value for a given state-action pair, and the immediate reward is denoted by $ r $. The discount factor $ \gamma $ is used to weigh future rewards, while the term $ \alpha \log \pi_{\theta}(a'|s') $ is the entropy term that encourages exploration by increasing the randomness of the action selection. The minimum over two Q-value networks $ \phi_1 $ and $ \phi_2 $ is taken to reduce the overestimation of the Q-values.

\subsubsection{Actor Policy Loss}

The actor policy loss function $ L_{\text{Actor}} $ is designed to maximize both the expected return and the entropy of the policy. The loss function is given by:

\begin{equation}
L_{\text{Actor}}(\theta) = \mathbb{E}_{s \sim \mathcal{D}}\left[ \alpha \log \pi_{\theta}(a|s) - Q_{\phi}(s, a) \right]
\end{equation}

This function combines the Q-value function $ Q_{\phi}(s, a) $ with the entropy term $ \alpha \log \pi_{\theta}(a|s) $. The goal is to maximize the expected return while ensuring that the policy remains sufficiently exploratory by maintaining a certain level of randomness in the action distribution.

\section{Proposed Methodology}

\subsection{Overview}

\begin{figure}[t]
    \centering
    \includegraphics[width=1\linewidth]{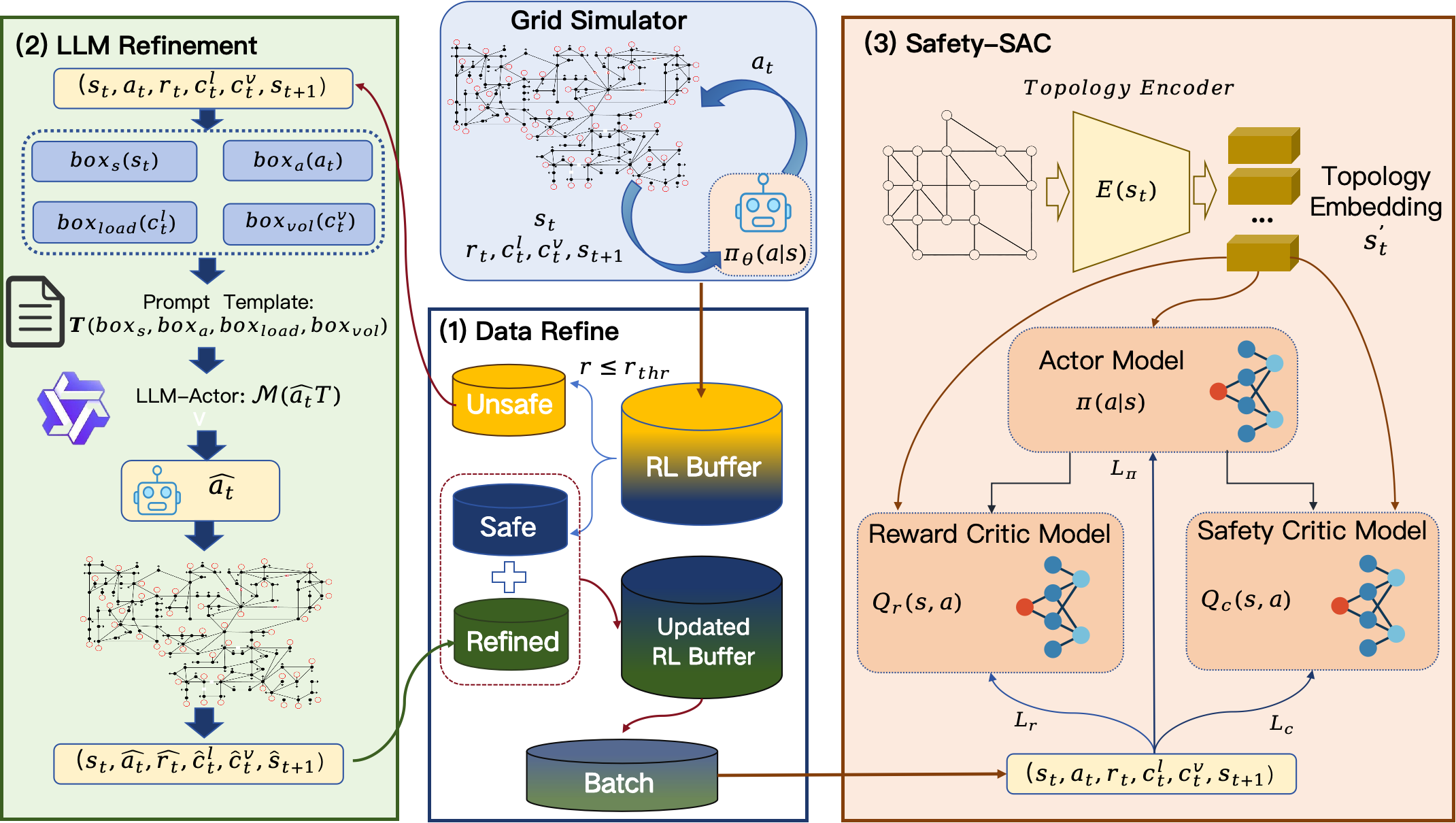}
    \caption{Method Overview.The framework first filters and refines transitions using a reward threshold, then leverages the LLM to correct unsafe samples, and finally trains the Safety-SAC agent on the updated buffer to improve both performance and safety.}
    \label{fig:framework}
\end{figure}

In this section, we present a novel approach that integrates Safety reinforcement learning with large language models (LLMs) for the safe and efficient operation of power systems, specifically targeting the Grid2OP environment, as shown as Fig.\ref{fig:framework}.  
{\color{red}Building upon the standard SAC formulation introduced in Sec.\ref{sec:2.2}, we extend the objective with safety-cost signals and introduce a dual-critic structure to handle operational constraints. Further, our methodology combines this safety-aware Soft Actor--Critic backbone with an LLM-based operator, so that the learned policy not only optimizes operational performance but also explicitly accounts for voltage and line-loading security constraints. }

{\color{red}The proposed framework operates at the experience level. Unlike policy-level knowledge injection or reward shaping, the LLM refines selected replay-buffer transitions without modifying the policy network or reward function. All refined actions are validated by the power-flow simulator before storage. This design preserves physical consistency and maintains compatibility with standard off-policy RL algorithms under discrete combinatorial topology actions.

Moreover, unlike methods that rely solely on limited experience or predefined domain rules, the LLM can draw upon knowledge from pre-training to inform its refinement, including general reasoning about system topology, load-flow patterns, and operational heuristics. This allows the framework to benefit from flexible reasoning that is not confined to hand-crafted rules, while the simulator validation ensures that all refined actions remain physically feasible.}
\subsubsection{Interaction with Environment}

We now describe the interaction between the actor (represented by the LLM-Actor and the RL policy) and the Grid2OP environment. The goal is to model the continuous loop in which the agent observes the system state, selects an action, receives the next state and feedback, and stores this information for subsequent learning.

\subsubsection{Action Selection by the Actor}

At each discrete time step $t$, the agent observes the current environment state $s_t$, which encodes the relevant grid information such as voltage magnitudes, line flows, and load levels. The actor then selects a control action $a_t$ according to a stochastic policy
\begin{equation}
  a_t = \pi_{\theta}(s_t),
\end{equation}
where $\pi_{\theta}$ denotes the policy function parameterized by $\theta$, and $a_t$ is the topological action applied to the Grid2OP environment at time $t$. In our framework, this policy is implemented by an LLM-Actor combined with the underlying Safety-SAC network, as detailed in the subsequent sections.

\subsubsection{Grid2OP Environment and State Transition}

Given the selected action $a_t$, the Grid2OP environment updates the grid topology accordingly, for example by opening or closing transmission lines or reconfiguring substations. The environment then computes the resulting power flows and voltages and produces the next state $s_{t+1}$. The state transition can be abstracted as
\begin{equation}
  s_{t+1} = \mathcal{T}(s_t, a_t),
\end{equation}
where $\mathcal{T}(\cdot)$ denotes the state-transition function determined by the Grid2OP simulator. This formulation emphasizes that the agent has no explicit model of $\mathcal{T}$ and instead learns solely from observed transitions.

\subsubsection{Voltage Violation and Line Overload Indicators}

Once the new state $s_{t+1}$ is obtained, the environment evaluates whether any security limits are violated. In this work, we focus on two types of constraints: voltage violations and line overloads.
\begin{itemize} \item Voltage Violation: The voltage at each node in the grid is computed, and a violation occurs if the voltage falls outside the allowable range, typically defined as ($V_{\text{min}}, V_{\text{max}}$). The percentage of nodes violating the voltage limits is computed as: 
\begin{equation} 
P_{\text{voltage}} = \frac{1}{N} \sum_{i=1}^{N} \mathbb{I}(V_i \notin [V_{\text{min}}, V_{\text{max}}]) 
\label{eq:P_voltage}
\end{equation} 
where $N$ is the total number of nodes, ($V_i$) is the voltage at node ($i$), and ($\mathbb{I}$) is the indicator function that equals 1 if ($V_i$) is outside the allowed range, and 0 otherwise. 
\item Line Overload: Similarly, the load on each transmission line is computed, and an overload occurs if the line’s flow exceeds its capacity. The overload ratio for each line is calculated as: 
\begin{equation} 
P_{\text{overload}} = \frac{1}{M} \sum_{j=1}^{M} \mathbb{I}(P_j > P_{\text{max},j}) 
\label{eq:P_overload}
\end{equation} 
where ($M$) is the number of lines, ($P_j$) is the flow on line ($j$), and ($P_{\text{max},j}$) is the maximum capacity of line ($j$). The indicator function ($\mathbb{I}$) equals 1 if the line flow ($P_j$) exceeds the line's maximum capacity, and 0 otherwise. 
\end{itemize}

For completeness, we denote by $C_{t}^{v}$ the fraction of buses whose voltages fall outside the admissible range at time $t$, and by $C_{t}^{l}$ the fraction of lines whose flows exceed their thermal limits at time $t$. These indicators serve as the basic ingredients for constructing the safety-cost signal and the safety objective $J_{\mathrm{safe}}(\tau)$ in \eqref{eq:J_safe}.

\subsubsection{Feedback and Replay Buffer Construction}

After applying $a_t$ and computing the next state $s_{t+1}$, the environment returns a performance reward $r_t$ and safety cost $C_{t}^{v}$ and $C_{t}^{l}$ associated with that transition. At time $t$ we construct the transition tuple
\begin{equation}
  \tau_t = 
  \bigl(
    s_t,\,
    a_t,\,
    r_t,\,
    C_{t}^{v},\,
    C_{t}^{l},\,
    s_{t+1}
  \bigr).
\end{equation}
Each new transition $\tau_t$ is then appended to the replay buffer $\mathcal{D}$, i.e.,
\begin{equation}
  \mathcal{D} \leftarrow \mathcal{D} \cup \{\tau_t\},
\end{equation}
so that $\mathcal{D}$ accumulates all observed transitions over time. The buffer $\mathcal{D}$ is later sampled to train the Safety-SAC agent.

\subsection{Problem Formulation Exchange}

The optimization problem in Section~\ref{sec:2.1} specifies the operational objective together with current and voltage limits. In the original formulation, the feasible set is strictly determined by the line-current constraint~\eqref{eq:current_limit} and bus-voltage bounds~\eqref{eq:voltage_limit}, both of which must be satisfied at every time step. Although these hard constraints are essential for system security, they heavily restrict exploration for an RL agent, especially during early policy learning. Exploratory actions that trigger any limit violation immediately push the system outside the feasible region, yielding only binary feasible/infeasible signals and providing no gradient information for improving the policy. This often results in conservative behavior and premature convergence to suboptimal solutions.

To overcome this limitation, we relax the hard-constraint treatment by recasting them as a \emph{safety objective}. Instead of enforcing~\eqref{eq:current_limit}--\eqref{eq:voltage_limit} pointwise, we use the violation indicators defined in~\eqref{eq:P_voltage}--\eqref{eq:P_overload}, which measure the fraction of buses with voltage deviations and the fraction of lines exceeding their thermal ratings at each time step. Using these indicators, we define a cumulative safety metric:
\begin{equation}
  J_{\mathrm{safe}}(\tau)
  = \frac{1}{n}\sum_{t=1}^{n}
  \big(
    \alpha_{\mathrm{v}} C_{t}^{v}
    + \alpha_{\mathrm{l}} C_{t}^{l}
  \big),
  \label{eq:J_safe}
\end{equation}
{\color{red}where $\alpha_{\mathrm{v}}>0$ and $\alpha_{\mathrm{l}}>0$ weight} the impact of voltage and overload violations, and $n$ is the rollout horizon. This reformulation converts strict feasibility into a smooth, informative safety-risk signal that is more compatible with RL training.

Consequently, the overall decision-making problem is no longer a single-objective constrained formulation, but rather a bi-objective optimization problem consisting of the operational objective defined previously, $J_{\mathrm{op}}(\tau)$, and the newly introduced safety objective, $J_{\mathrm{safe}}(\tau)$:
\begin{equation}
  \min_{\tau}\; J_{\mathrm{op}}(\tau),
  \qquad
  \min_{\tau}\; J_{\mathrm{safe}}(\tau).
  \label{eq:biobjective}
\end{equation}
In practice, these two objectives may be scalarized into a single surrogate objective,
\begin{equation}
  J(\tau) = J_{\mathrm{op}}(\tau) 
  + \kappa\, J_{\mathrm{safe}}(\tau),
  \label{eq:scalarization}
\end{equation}
where $\kappa \ge 0$ controls the safety–performance trade-off. This formulation preserves the original economic objective while explicitly embedding safety as a second minimization target. More importantly, it provides a smooth optimization structure that is compatible with gradient-based reinforcement-learning algorithms and forms the foundation of our Safety Soft Actor–Critic framework.

\subsection{Safety Soft Actor-Critic Method}

Building on the bi-objective formulation in \eqref{eq:biobjective}--\eqref{eq:scalarization}, we now instantiate a concrete reinforcement-learning algorithm that can handle both operational performance and safety requirements. As shown as Algorithm\ref{alg:safety_sac}, we adopt a Safety Soft Actor--Critic (Safety-SAC) structure, in which the operational objective $J_{\mathrm{op}}(\tau)$ and the safety objective $J_{\mathrm{safe}}(\tau)$ are represented by a reward signal and a safety-cost signal, respectively. These signals are then used to train a dual-critic architecture with a shared encoder, and an actor that is optimized under both the entropy-regularized reward objective and a safety-penalization objective. The resulting framework naturally fits within the constrained Markov decision process (CMDP) paradigm and provides a principled way to trade off efficiency and risk.

\subsubsection{Reward and Safety Cost}

Based on the soft-constrained objective, we decompose the learning signal into an operational reward and a safety cost. The instantaneous reward $r_t$ is constructed to mirror the economic and tracking objectives in~\eqref{eq:obj}. Specifically, at each time step $t$ we define
\begin{equation}
  r_t = 
  - \sum_{p\in\mathcal{E}}\Big(\frac{L_{p,t}}{P_{p,t}}\Big)
  - \text{penalty}\cdot\Delta t \cdot\big(P_{p,t}-P^{\mathrm{ref}}_{p,t}\big)^{2},
  \label{eq:reward}
\end{equation}
so that higher efficiency and closer tracking of the reference injections $P^{\mathrm{ref}}_{p,t}$ correspond to larger reward values.

In parallel, we introduce a safety-cost signal $C_t$ to quantify the risk associated with constraint violations. Using the violation ratios defined in~\eqref{eq:P_voltage} and~\eqref{eq:P_overload}, the safety cost at time $t$ is given by
\begin{equation}
  C_t = 
  \alpha_{\mathrm{v}}\,C_{t}^{v}
  + \alpha_{\mathrm{l}}\,C_{t}^{l},
  \label{eq:safety_cost}
\end{equation}

The pair $(r_t,C_t)$ thus provides a dual learning signal: $r_t$ encourages operation that is economically efficient and close to the nominal schedule, whereas $C_t$ penalizes actions that drive the system toward unsafe regimes. Within the constrained Markov decision process (CMDP) framework, the control objective is to maximize the expected discounted return of $\{r_t\}$ subject to an upper bound on the expected discounted accumulation of $\{C_t\}$.

\subsubsection{Safety Constraint Critic}

To perform safety-aware value estimation, we employ a dual-critic architecture built upon a shared state encoder. Let $E(\cdot)$ be an encoder network mapping state $s_t$ to a latent representation
\begin{equation}
  s'_t = E(s_t).
  \label{eq:encoder}
\end{equation}
The latent state $s'_t$ is then used by two distinct critics: a reward critic $Q_r(s',a)$ and a safety critic $Q_c(s',a)$.

The reward critic estimates the entropy-regularized action-value function of the SAC objective. Given a transition $(s_t,a_t,r_t,C_t,s_{t+1})$ from the replay buffer, we first obtain $s'_t$ and $s'_{t+1}$ via~\eqref{eq:encoder}. The target for $Q_r$ is defined as
\begin{equation}
  y_r = r_t 
  + \gamma\,\mathbb{E}_{a'\sim\pi(\cdot|s'_{t+1})}
  \big[ Q_r(s'_{t+1},a') - \alpha \log \pi(a'|s'_{t+1}) \big],
  \label{eq:yr}
\end{equation}
where $\gamma\in(0,1)$ is the discount factor and $\alpha>0$ is the temperature parameter controlling the strength of entropy regularization. The reward-critic loss is then
\begin{equation}
  L_r = \mathbb{E}\big[(Q_r(s'_t,a_t) - y_r)^2\big].
  \label{eq:Lr}
\end{equation}

The safety critic $Q_c$ approximates the discounted cumulative safety cost. Its temporal-difference target is given by
\begin{equation}
  y_c = C_t 
  + \gamma\,\mathbb{E}_{a'\sim\pi(\cdot|s'_{t+1})}
  \big[ Q_c(s'_{t+1},a') \big],
  \label{eq:yc}
\end{equation}
leading to the loss
\begin{equation}
  L_c = \mathbb{E}\big[(Q_c(s'_t,a_t) - y_c)^2\big].
  \label{eq:Lc}
\end{equation}
Because both critics share the encoder $E(\cdot)$, the encoder parameters must capture features that are simultaneously informative for reward prediction and safety-cost estimation. To coordinate these learning objectives, we employ a Lagrangian-style joint loss for updating the encoder and critic networks,
\begin{equation}
  L_{\mathrm{enc}} = L_r + \lambda_c L_c,
  \label{eq:Lenc}
\end{equation}
where $\lambda_c\geq 0$ is a Lagrange multiplier regulating the emphasis on safety-value accuracy. By minimizing~\eqref{eq:Lenc}, the training procedure encourages the latent representation to embed both performance-related and safety-related information, which is crucial for stable safety-constrained policy improvement.

\subsubsection{Safety Constraint Actor}

The actor aims to learn a stochastic policy $\pi(a|s')$ that maximizes long-term reward while enforcing the safety constraint in expectation. Inspired by SAC, the reward-driven component of the actor objective is
\begin{equation}
  L_{\pi,r} = 
  \mathbb{E}_{a\sim\pi(\cdot|s')}
  \big[\alpha\log\pi(a|s') - Q_r(s',a)\big],
  \label{eq:Lpir}
\end{equation}
which encourages actions with high soft $Q_r$-values and high entropy.

To incorporate safety considerations, we use the safety critic $Q_c$ to evaluate the expected safety cost of candidate actions and introduce a margin-based penalty relative to a safety threshold $\epsilon_c$. The safety-driven component of the actor loss is defined as
\begin{equation}
  L_{\pi,c} = 
  \mathbb{E}_{a\sim\pi(\cdot|s')}
  \big[ \max\big(0, Q_c(s',a) - \epsilon_c\big) \big],
  \label{eq:Lpic}
\end{equation}
which is consistent with the constrained-optimization surrogate
\begin{equation}
  L_{\mathrm{co}} = 
  \max\big(0, \mathbb{E}_{(s',a)\sim\pi}[Q_c(s',a)] - \epsilon_c \big).
  \label{eq:Lco}
\end{equation}
Equations~\eqref{eq:Lpic}–\eqref{eq:Lco} penalize the policy whenever the predicted safety cost exceeds the tolerated level $\epsilon_c$, thereby steering the policy away from potentially hazardous control actions.

The overall actor objective combines the reward and safety terms as
\begin{equation}
  L_{\pi} = L_{\pi,r} + \beta L_{\pi,c},
  \label{eq:Lpi}
\end{equation}
where $\beta>0$ is a trade-off coefficient. Minimizing~\eqref{eq:Lpi} yields a Safety Soft Actor--Critic (Safety-SAC) policy that simultaneously exploits the reward critic for performance improvement and the safety critic for risk mitigation.

\begin{algorithm}
\small
\caption{Safety Soft Actor--Critic (Safety-SAC)}
\label{alg:safety_sac}

\DontPrintSemicolon

\KwIn{Environment, encoder $E$, discount $\gamma$, temperature $\alpha$, 
safety weight $\beta$, learning rates $\eta_\pi,\eta_Q,\eta_\lambda$.}
\KwOut{Safety-aware policy parameters $\theta_\pi$.}

Initialize encoder $\theta_E$, reward critic $\theta_r$, safety critic $\theta_c$, actor $\theta_\pi$\;
Initialize target critics $\bar{\theta}_r \leftarrow \theta_r$, $\bar{\theta}_c \leftarrow \theta_c$\;
Initialize Lagrange multiplier $\lambda_c \ge 0$ and replay buffer $\mathcal{D}$\;

\For{each episode}{
  Receive initial state $s_0$, set $t \leftarrow 0$\;
  \While{episode not terminated}{
    Encode $s'_t = E_{\theta_E}(s_t)$, sample action $a_t \sim \pi_{\theta_\pi}(\cdot|s'_t)$\;
    Apply $a_t$, observe $s_{t+1}$, compute $r_t$ via \eqref{eq:reward} and $C_t$ via \eqref{eq:safety_cost}\;
    Store $(s_t,a_t,r_t,C_t,s_{t+1})$ in $\mathcal{D}$, set $t \leftarrow t+1$\;

    \For{each gradient step}{
      Sample minibatch $\{(s_i,a_i,r_i,c_i,x'_i)\}_{i=1}^{B}$ from $\mathcal{D}$\;
      Encode $s'_i = E(s_i)$, $s''_i = E(x'_i)$, sample $a'_i \sim \pi(\cdot|s''_i)$\;

      \tcp{Critic and encoder update}
      Compute $y_{r,i}$ via \eqref{eq:yr} and $y_{c,i}$ via \eqref{eq:yc}\;
      Compute critic losses $L_r$ and $L_c$ via \eqref{eq:Lr}, \eqref{eq:Lc}\;
      Set $L_{\mathrm{enc}} = L_r + \lambda_c L_c$ and update 
      $\theta_E,\theta_r,\theta_c \leftarrow \theta_E,\theta_r,\theta_c 
      - \eta_Q \nabla_{\theta_E,\theta_r,\theta_c} L_{\mathrm{enc}}$\;

      \tcp{Lagrange multiplier update}
      Estimate $\widehat{L}_{\mathrm{co}} = \max\bigl(0, \tfrac{1}{B}\sum_i Q_c(s'_i,a_i) - \epsilon_c\bigr)$\;
      Update $\lambda_c \leftarrow [\lambda_c + \eta_\lambda \widehat{L}_{\mathrm{co}}]_+$\;

      \tcp{Actor update}
      Sample $\tilde{a}_i \sim \pi_{\theta_\pi}(\cdot|s'_i)$, compute $L_{\pi,r}$ and $L_{\pi,c}$ via 
      \eqref{eq:Lpir}, \eqref{eq:Lpic}\;
      Form $L_\pi = L_{\pi,r} + \beta L_{\pi,c}$ and update 
      $\theta_\pi \leftarrow \theta_\pi - \eta_\pi \nabla_{\theta_\pi} L_\pi$\;

      \tcp{Target networks soft update}
      $\bar{\theta}_r \leftarrow \tau\theta_r + (1-\tau)\bar{\theta}_r$,
      $\bar{\theta}_c \leftarrow \tau\theta_c + (1-\tau)\bar{\theta}_c$\;
    }
  }
}
\end{algorithm}

\subsection{Large Language Model Operator}

In this section, we present the Large Language Model (LLM) Operator, which acts as a knowledge-based decision module on top of the Safety-SAC backbone. The LLM-Actor is responsible for interpreting and refining the actions selected by the reinforcement-learning agent based on historical experience stored in the replay buffer. The frequency of invoking the LLM to refine the buffer is a tunable hyperparameter $f$, representing how many training iterations of Safety-SAC are performed before each LLM-based buffer refinement. By utilizing a pre-trained large language model together with a structured prompt template, the LLM-Operator analyses the current state, past actions, and safety indicators, and proposes modified actions that are more consistent with safe and efficient operation.

\subsubsection{Pre-trained LLM as Actor}
\label{sec:llm_Actor}
The LLM-Actor uses a pre-trained LLM directly for power-system decision-making. At selected time steps, the LLM examines transitions in the replay buffer that exhibit poor performance. For each stored transition
\[
  (s_t, a_t, r_t, C_{t}^{v}, C_{t}^{l}, s_{t+1}) \in \mathcal{D},
\]
we identify those whose reward is below a prescribed threshold $r_{thr}$,
\begin{equation}
  r_t < r_{thr}.
\end{equation}
These transitions correspond to suboptimal actions that reduce economic efficiency or increase safety risks. The LLM-Actor then analyzes these cases and proposes alternative actions to improve both reward and safety, using the LLM’s inherent knowledge about power-system behavior.

\newcommand{\var}[1]{\textcolor{blue}{\texttt{\{#1\}}}}

\begin{figure}
\centering

\begin{tcolorbox}[
    colback=white,      
    colframe=black,     
    arc=2pt,            
    boxrule=0.6pt,      
    width=0.97\linewidth
]

\footnotesize
\ttfamily

You are an expert power grid operator with deep knowledge of
transmission line overloads and voltage regulation.\medskip

NOW, let's analyze the current situation at \var{timestamp} step by step.\medskip

1. Grid overview \\
\quad -- Total elements: \var{num\_elements} \\
\quad -- Operable substations: \var{controllable\_substations} \\
\quad -- Total lines: \var{num\_lines} \\
\quad -- Voltage normal range: \var{v\_low\_threshold}--\var{v\_high\_threshold} pu\medskip

2. Top-\(K\) overloaded lines (threshold: \var{overload\_threshold}\%) \\
\quad For each overloaded line \(L\): \\
\quad -- Line \var{line\_id} (\var{usage\_percentage}\%, \var{severity}) \\
\quad -- Connection: Sub \var{or\_sub} \(\leftrightarrow\) Sub \var{ex\_sub} \\
\quad -- Active power flow: \var{p\_or} MW\medskip

3. Voltage abnormalities \\
\quad Under-voltage nodes (<\var{v\_low\_threshold} pu): Node \var{node\_id}: \var{v\_node} pu \\
\quad Over-voltage nodes (>\var{v\_high\_threshold} pu): Node \var{node\_id}: \var{v\_node} pu\medskip

4. Crucial substations (overload / voltage-related) \\
\quad -- Bus 0 lines: \var{bus0\_lines} \\
\quad -- Bus 1 lines: \var{bus1\_lines} \\
\quad -- Disconnected lines: \var{disconnected\_lines} \\
\quad -- Voltage nodes: \var{node\_id}: \var{v\_node} pu\medskip

5. Bad action examples (RL-derived) \\
\quad -- Avoid: \var{line\_id : new\_bus\_id} \\
\quad -- Reward: \var{reward} \\
\quad -- Voltage impact: \var{description\_of\_voltage\_change}\medskip

6. Operational constraints \\
\quad -- Lines in cooldown: \var{cooldown\_lines} \\
\quad -- Remaining steps: \var{cooldown\_steps}\medskip

Please provide your response in the following format: \\
1. [Analysis of critical issues] \\
2. [Analysis of current topology] \\
3. [Deep reasoning of bad line change examples] \\
4. Proposed line changes (bus\_id must be 0 or 1)\medskip

proposed LINE changes: \var{line\_id : new\_bus\_id}

\end{tcolorbox}

\caption{Prompt template used for power grid control.
\textcolor{blue}{Blue-colored tokens} denote variables extracted from the replay buffer at timestep \(D_t\), including observations, overload indicators, voltage measurements, and RL-derived action feedback. }
\label{fig:prompt-template}
\end{figure}

\subsubsection{Action Prompt Construction}

To apply the LLM for safety analysis, we convert each selected transition into a textual prompt containing the system state, the action taken, the reward, and the safety indicators. The prompt is generated through a template function that formats these elements into a structured string, as shown as Fig.\ref{fig:prompt-template}. This standardized template ensures that the LLM receives consistent and sufficient context to evaluate the transition and provide meaningful refinements. {\color{red}The structured template explicitly exposes safety-critical signals and operational constraints to the LLM. This design improves reasoning consistency and ensures that refined actions remain compliant with physical and operational limits. The LLM output follows a fixed schema with a machine-parseable action list \(\{\texttt{line\_id}:\texttt{new\_bus\_id}\}\), where \(\texttt{new\_bus\_id}\in\{0,1\}\), enabling deterministic parsing and simulator-based validation.}

In this approach, the prompt at each time step $t$ is formulated as:
\begin{equation}
\mathrm{Prompt}_t
= \mathrm{T}\bigl(
s_t,
a_t,
r_t,
C_{t}^{v},
C_{t}^{l}
\bigr),
\end{equation}
where $s_t$ represents the state, $a_t$ is the action taken at time $t$, $r_t$ is the corresponding reward, and $C_{t}^{v}$ and $C_{t}^{l}$ are the voltage-violation and line-overload ratios, respectively.











\subsubsection{LLM-Based Action Refinement}

Given the constructed prompt, the pre-trained LLM generates a response that reflects its understanding of the underlying power-system dynamics and operational heuristics. Instead of learning a full Markovian control policy, the LLM focuses on refining individual actions based on the current transition and its safety/performance attributes. Let the refined action proposed by the LLM be denoted by $a_{\mathrm{LLM}}$. We abstract this mapping as
\begin{equation}
  a^{\mathrm{LLM}}_{t} 
  = \mathcal{M}\bigl(
      \mathrm{Prompt}_t
    \bigr),
\end{equation}
where $\mathcal{M}(\cdot)$ represents the LLM-based modification function. The output $ a^{\mathrm{LLM}}_{t}$ is a refined action suggestion that is expected to reduce the likelihood of voltage violations and line overloads while improving the overall reward.

Since the output of LLM is typically textual, we introduce an extraction function $\mathcal{E}(\cdot)$ that parses the LLM response and converts it into a valid action vector compatible with the Grid2OP interface and the Safety-SAC policy. The final action supplied to the RL agent is given by
\begin{equation}
  \bar{a_{t}} = \mathcal{E}( a^{\mathrm{LLM}}_{t}),
\end{equation}
where $\bar{a_{t}}$ has the same format as actions in the original policy space (e.g., a structured topology-change vector). 

By incorporating this LLM-based refinement mechanism, the framework expands the agent’s action space toward safer decisions and significantly improves its overall performance.

\subsection{\textcolor{blue}{Mechanism of LLM-Based Transition Refinement}}
\label{sec:llm_refinement_mechanism}

\textcolor{blue}{This subsection clarifies the functional role of the Large Language Model (LLM)
in the proposed framework. The LLM operates exclusively at the replay-buffer level and complements the Safety-SAC backbone by selectively refining historically observed transitions, without modifying the learning algorithm or the environment dynamics.}{\color{blue}As introduced in Section~\ref{sec:llm_Actor}, LLM-based refinement is triggered only for transitions associated with low performance or elevated risk. For a transition}
\begin{equation}
\tau_t = \left( s_t, a_t, r_t, C_{t}^{v},C_{t}^{l}, s_{t+1} \right),
\end{equation}
\textcolor{blue}{refinement is applied when the reward satisfies $r_t < r_{\mathrm{thr}}$, a criterion already defined in the LLM operator module. Transitions meeting this condition typically correspond to unsafe or inefficient topology actions. Restricting LLM intervention to this subset avoids unnecessary modification of high-quality experiences and preserves replay-buffer diversity.}

\textcolor{blue}{For each selected transition, the state $s_t$, action $a_t$, and safety indicators are converted into a structured prompt. Based on this information, the LLM proposes a refined action $\widehat{a}_t$ using contextual reasoning grounded in power-system operational knowledge. Importantly, the LLM does not predict the next state or reward. The refined action is validated
and executed in the Grid2Op simulator to obtain the true system transition}
\begin{equation}
s_{t+1} = Grid2Op\left( s_t, \widehat{a}_t \right).
\end{equation}

\textcolor{blue}{The resulting transition $\widehat{\tau}_t = \left( s_t, \widehat{a}_t, \widehat{r}_t, \widehat{C}_{t}^{v},\widehat{C}_{t}^{l}, \widehat{s}_{t+1} \right)$ is appended to the replay buffer alongside the original sample, ensuring that all learning signals remain physically consistent.}

\begin{figure}
    \centering
    \includegraphics[width=1\linewidth]{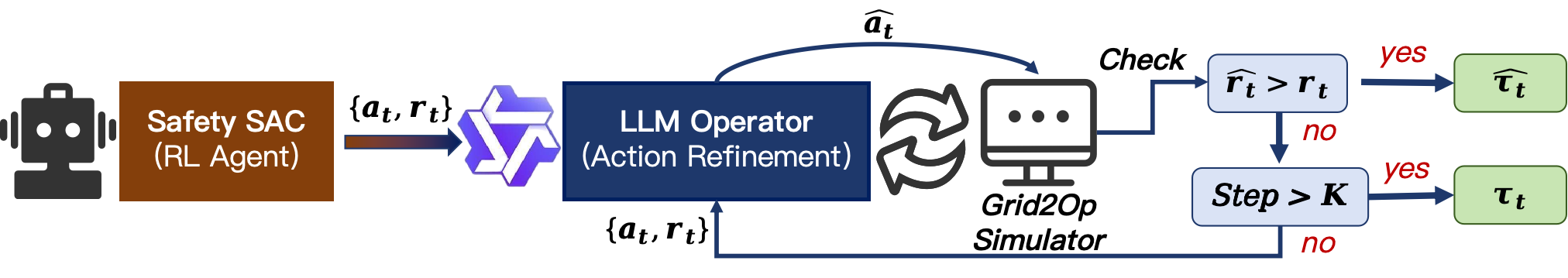}
    \caption{LLM-based transition refinement mechanism. 
The Safety-SAC agent generates candidate actions, which are selectively refined by the LLM under reward-based triggering conditions.Parsing and feasibility checks ensure that only valid actions are executed.}
    \label{fig:llm_refinement}
\end{figure}

{\color{red}The LLM output is parsed and validated before execution, as indicated in the Fig.~\ref{fig:llm_refinement}. If no improved action is found after modifying the number of steps $K$ (set to 3 in the experimental setup), the original action should be output. Invalid actions are discarded and do not alter the original transition.}

\textcolor{blue}{From a learning perspective, the LLM-based mechanism can be viewed as targeted experience curation rather than global data redistribution. By augmenting the buffer with safer alternatives to historically poor actions, policy updates are biased toward lower-risk regions of the action space. This suppresses unsafe exploration early in training and leads to smoother convergence and reduced oscillations, consistent with the training dynamics reported in Sections~\ref{sec:trends36} and~\ref{sec:trends118}.}

\section{Experiment}
\label{sec:experiment}
\subsection{Environmental Setup}
\begin{figure}[t]
    \centering
    \includegraphics[width=0.8\linewidth]{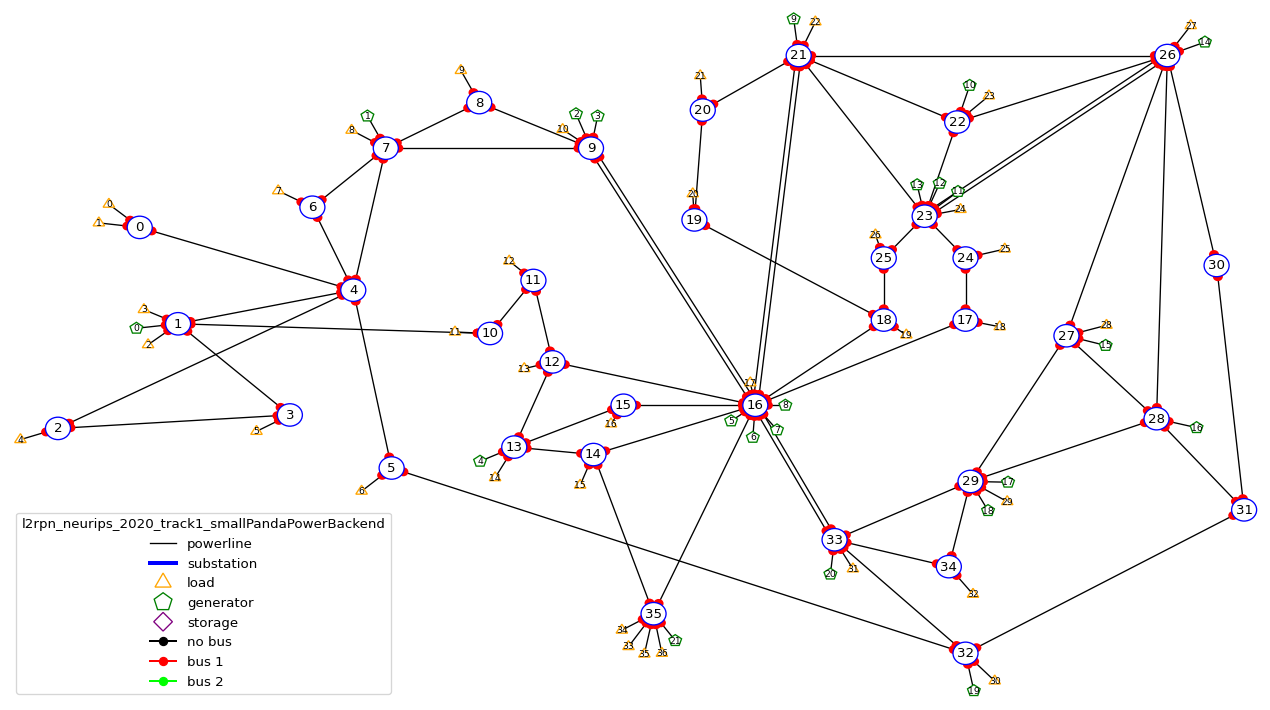}
    \caption{Topology diagram of the IEEE 36-bus system.}
    \label{fig:36case}
\end{figure}

\begin{figure}[t]
    \centering
    \includegraphics[width=0.8\linewidth]{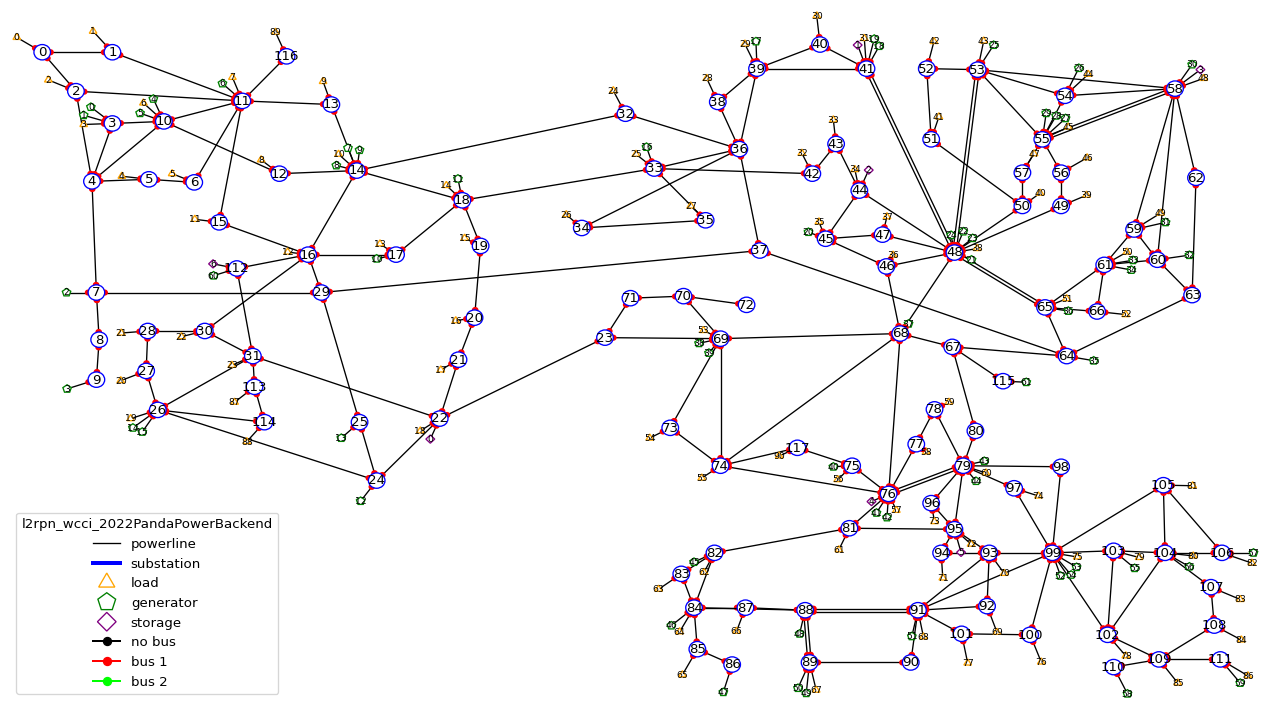}
    \caption{Topology diagram of the IEEE 118-bus system.}
    \label{fig:118case}
\end{figure}
Performance validation employs two multi-scale benchmark environments provided by the Grid2Op platform \cite{grid2op}. The 36-bus system in Fig.\ref{fig:36case}) originates from the L2RPN WCCI 2020 competition \cite{marot2021learning}, replicating operational characteristics of the US Midwest transmission corridor with 36 substations, 59 branches, and 22 generation units. This configuration presents $6 \times 10^4$ combinatorial switching decisions while capturing stochastic load variations and scheduled outage events. Testing utilizes 10 episodes of 864-step (72-hour) trajectories spanning multiple difficulty tiers. Scalability evaluation leverages the IEEE 118-bus synthetic system (Fig.\ref{fig:118case}) from the L2RPN WCCI 2022 competition \cite{marot2022learning}, comprising 186 power lines, 91 loads, and 62 generators with elevated renewable integration. The expanded topology yields $7 \times 10^4$ reconfiguration alternatives, intensifying combinatorial complexity under generation uncertainty. Training data encompasses 32-year chronological sequences at 5-minute granularity, totaling 1.7 GB across $3.36 \times 10^6$ temporal snapshots.

\subsection{Comparative Metric}

To quantitatively evaluate the performance of reinforcement learning (RL) agents for power system control, we adopt a set of metrics widely used in power system operation and reinforcement learning. All metrics are computed across multiple rollouts and averaged to reduce stochastic variability.

\begin{itemize}
    \item \textbf{Survival Step}: The number of simulation steps completed before an episode is terminated due to system instability or collapse. Higher values indicate improved grid reliability and the agent's capability to maintain secure operation over longer horizons.

    \item \textbf{Reward}: The cumulative reward accumulated along a rollout, defined as $\sum_{t=0}^{T} r_t$. This metric captures overall control performance, reflecting operational objectives, system reliability incentives, and penalty terms. Since early termination reduces reward accumulation, higher rewards generally imply more robust and effective control policies.

    \item \textbf{Overload Rate}: The percentage of time steps during which at least one transmission line exceeds its thermal rating. Lower overload rates indicate that the agent effectively mitigates congestion and alleviates stress on critical network components, thus reducing the risk of cascading failures.

    \item \textbf{Voltage Violation Rate}: The percentage of time steps in which at least one bus voltage magnitude deviates from its acceptable operating limits. Lower violation rates indicate improved voltage regulation and enhanced compliance with standard grid operating constraints.

    \item \textbf{SafetyCost}: A composite safety indicator defined as
    \begin{equation}
        \mathrm{SafetyCost} = 0.9\times\mathrm{OverloadRate} + 0.1\times\mathrm{ViolationRate}.
    \end{equation}
    Lower SafetyCost values represent safer and more constraint-aware decision-making by the RL agent. {\color{red}\textbf{Overload Rate} and \textbf{Voltage Violation Rate}  directly correspond to line thermal limit violations and bus voltage constraint violations in practical dispatch. SafetyCost aggregates these violations as a smooth learning surrogate while preserving their physical meaning.}
\end{itemize}

\subsection{Implementation Details}

{\color{red}The reward threshold is set to \( r_{\mathrm{thr}} = 0.3 \) to determine when LLM-based refinement is triggered. }{\color{blue}and the weighting coefficients for voltage and line-loading violations are chosen as \( \alpha_{v} = 0.9 \) and \( \alpha_{l} = 0.1 \). For the reinforcement learning module, we use a learning rate of \( 5\times10^{-5} \) and a batch size of 32; the agent is trained for 10{,}000 steps on a single NVIDIA A100 GPU, and the LLM is invoked every \(f=200\) steps to augment the replay buffer. The RL components \( E(\cdot) \), \( \pi(\cdot) \), and \( Q_{r}(\cdot) \) follows the ACE setup described in \cite{yoon2021winning,wan2025think}, and \(Q_{c}(\cdot)\) adopts the same backbone as \(Q_{r}(\cdot)\) with a sigmoid output to ensure values in $(0, 1)$; {\color{red}the tolerated level is set to \(\epsilon_c=1\) according to the normalized range of the safety critic output, activating penalties near the violation boundary.} Key model and training hyperparameters are summarized in Table~\ref{tab:hyperparams}.}

{\color{red}The proposed refinement mechanism is independent of a specific LLM architecture. We adopt a representative open-source model (Qwen-2.5-7B) to validate the framework, as the focus of this study is the interaction mechanism rather than model scaling effects.

Incorporating the LLM increases the training-stage computational requirement. In our experimental setup, GPU memory demand increases from approximately 16 GB (RL-only) up to approximately 80 GB when LLM refinement is enabled, and the training time extends from about 5 hours to 35 hours. This additional cost is incurred during training only, while real-time deployment does not require LLM inference.}
\begin{table}
  \centering
  \caption{Hyperparameter configuration used in experiments.}
  \begin{tabular}{lcc}
    \toprule
    Hyperparameter & Symbol & Value \\
    \midrule
    \multicolumn{3}{c}{\textbf{Training setup}} \\
    \midrule
    Total interaction steps & \(N_{\mathrm{int}}\) & \(10000\) \\
    Replay buffer capacity & \(|\mathcal{D}|\) & \(5000\) \\
    Max episode length & \(T_{\max}\) & \(864/2016\) \\
    \midrule
    \multicolumn{3}{c}{\textbf{RL optimization}} \\
    \midrule
    Batch size & \(B\) & \(32\) \\
    Actor learning rate & \(\eta_{\pi}\) & \(5\times10^{-5}\) \\
    Critic learning rate & \(\eta_{Q}\) & \(10^{-4}\) \\
    Encoder/embedding learning rate & \(\eta_{E}\) & \(5\times10^{-5}\) \\
    \midrule
    \multicolumn{3}{c}{\textbf{RL Model}} \\
    \midrule
    Attention head number & \(H\) & \(8\) \\
    Latent/embedding dimension & \(d\) & \(128\) \\
    Frame-stack length & \(n_{\mathrm{hist}}\) & \(6\) \\
    Dropout rate & \(p_{\mathrm{drop}}\) & \(0.1\) \\
    \midrule
    \multicolumn{3}{c}{\textbf{LLM}} \\
    \midrule
    LLM model & \(\mathcal{M}\) & Qwen-2.5-7B \\
    Max LLM-guided samples & \(N_{\mathrm{LLM}}\) & \(512\) \\
    LLM-Actor frequency & \(f_{\mathrm{actor}}\) & \(200\) \\
    Bad-sample reward threshold & \(r_{\mathrm{thr}}\) & \(0.3\) \\
    \bottomrule
  \end{tabular}
  \label{tab:hyperparams}
\end{table}

\subsection{Comparison Algorithms}

To benchmark the proposed framework, we compare four representative algorithms for topology reconfiguration.

\begin{itemize}
\item \textbf{SAC Algorithm\cite{yoon2021winning}.}
SAC serves as the baseline RL method, offering stable stochastic exploration but lacking mechanisms to enforce system-level safety constraints.

\item \textbf{Physics-guided RL (PGRL)\cite{dwivedi2024blackout}.} \textcolor{blue}{This baseline replaces the LLM-based module with a physics-guided heuristic while retaining the same RL backbone. For unsafe grid states, a small set of physically plausible corrective actions is constructed based on line loading ratios. Each candidate action is evaluated via one-step look-ahead power-flow simulation, and the action that most effectively alleviates overload without causing terminal failure is selected. The resulting transitions are used to guide learning, without relying on any language model.}

\item \textbf{ACE Algorithm\cite{wan2025think}.}
ACE augments RL with LLM-guided action suggestions. However, its LLM does not incorporate explicit safety considerations, making constraint violations still likely.

\item \textbf{ACE with Safety-SAC.}
This variant retains the ACE framework while replacing its SAC algorithm with our Safety-SAC, enabling safety-constraint guidance without modifying other ACE components.

\item \textbf{Ours.}
Our method combines Safety-SAC with a Safety LLM that filters or refines transitions using domain knowledge. This design jointly improves reward performance and adherence to safety constraints.

\end{itemize}

\begin{table}
  \centering
  \caption{The RL episode reward and survival step in IEEE 36-bus case.}
  \begin{tabular}{lcc}
    \toprule
    Method & Reward $\times10^{4}$ $\uparrow$ & Survival Step $\uparrow$  \\
    \midrule
    PGRL&1.570&513.7\\
    SAC & 1.643 & 518.3 \\
    ACE & 1.914 & 618.9 \\
    ACE with Safety-SAC & 2.162 & 683.9 \\
    Ours & \textbf{3.088} & \textbf{1028} \\
    \bottomrule
  \end{tabular}
  \label{tab:nips_reward}
\end{table}

\begin{table}
  \centering
  \caption{Comparative analysis of safety metrics in the IEEE 36-bus case.}
  \begin{tabular}{lccc}
    \toprule
    Method & Overload (\%) $\downarrow$ & Voltage (\%) $\downarrow$ & SafeCost $\downarrow$ \\
    \midrule
    PGRL&\textbf{0.451}&3.986&\textbf{0.522}\\
    SAC & 2.486 & 7.073 & 2.578 \\
    ACE & 1.379 & 2.463 & 1.400 \\
    ACE with Safety-SAC & 1.496 & 3.666 & 1.713 \\
    Ours & 0.817 & \textbf{1.441} & 0.880 \\
    \bottomrule
  \end{tabular}
  \label{tab:nips_safety}
\end{table}
\subsection{Results on IEEE 36-bus System}

Tables~\ref{tab:nips_reward} and~\ref{tab:nips_safety} summarize the quantitative evaluation on the IEEE 36-bus system. 
The following analysis integrates performance, safety, and training behaviors in a concise manner.
Across all metrics, the proposed method achieves the highest cumulative reward and the longest survival duration.
\textcolor{blue}{The physics-guided RL (PGRL) baseline demonstrates that physically motivated heuristics alone can already improve safety-related metrics compared with vanilla SAC, confirming the effectiveness of physics-guided experience refinement.}
While ACE benefits from the incorporation of Safety-SAC, introducing the Safety LLM brings a substantially larger improvement, indicating that LLM-guided refinement effectively corrects suboptimal and high-risk transitions.
The clear performance gap between our full model and the ``ACE with Safety-SAC'' variant further underscores the importance of knowledge-based reasoning in addition to low-level safety critics.


\begin{figure}[t]
    \centering
    \includegraphics[width=0.6\linewidth]{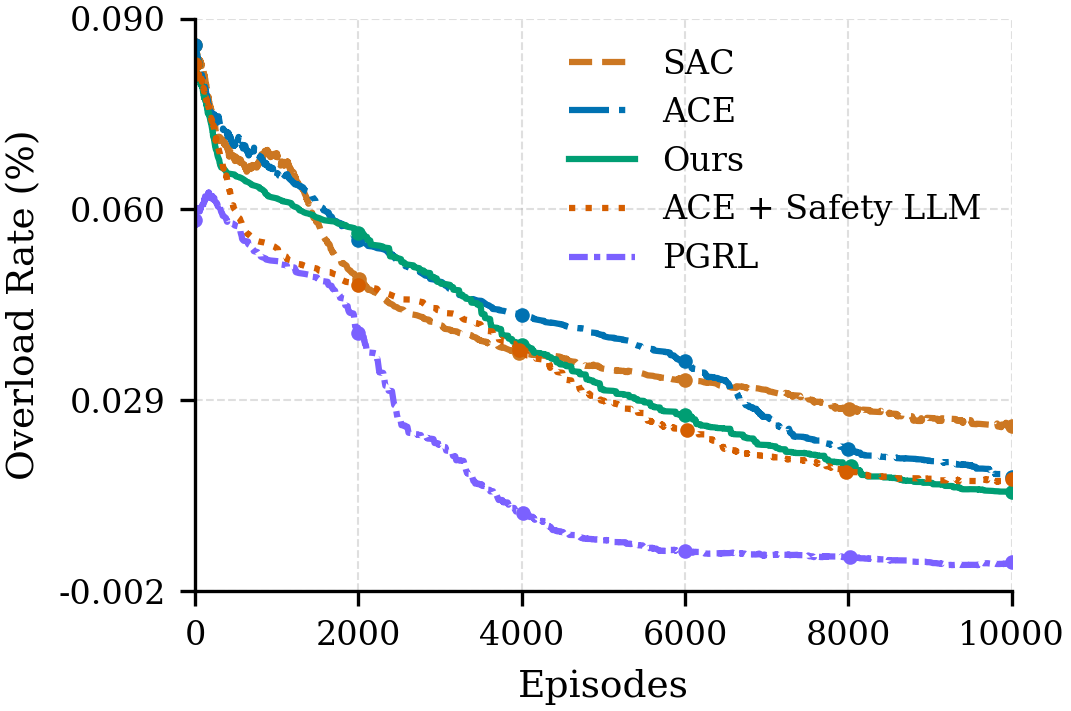}
    \caption{Overload rate of the IEEE 36-bus system.}
    \label{fig:36bus-overload}
\end{figure}

The safety evaluation results on the IEEE 36-bus system are presented in Table~\ref{tab:nips_safety}.
Overall, the proposed method achieves strong safety performance across multiple metrics.
It attains the lowest voltage violation rate (1.441\%) and a low SafeCost (0.880), indicating reliable and risk-aware operation.
\textcolor{blue}{Notably, the PGRL achieves the lowest overload rate and the minimum SafeCost, demonstrating that purely physics-based heuristics are highly effective in mitigating immediate overload risks.}
\textcolor{blue}{However, as shown in Table~\ref{tab:nips_reward}, this safety improvement comes at the expense of substantially lower cumulative reward and survival duration, reflecting limited adaptability and poor long-horizon performance.}

Among the baseline methods, ACE reduces safety violations compared with SAC, demonstrating the benefit of stronger action policies.
However, ACE with Safety-SAC does not consistently improve upon ACE and even increases violation rates in some cases, suggesting that low-level safety critics alone are insufficient for stable risk mitigation.
In contrast, our method effectively integrates LLM-based reasoning with safety-constraint reinforcement learning.
\textcolor{blue}{This integration enables a more favorable trade-off between safety and performance, achieving competitive safety levels while significantly improving survival duration and long-term reward.}
The substantial performance margin over learning-based baselines confirms the robustness and practicality of the proposed framework in complex distribution networks.

\begin{figure}[t]
    \centering
    \includegraphics[width=0.6\linewidth]{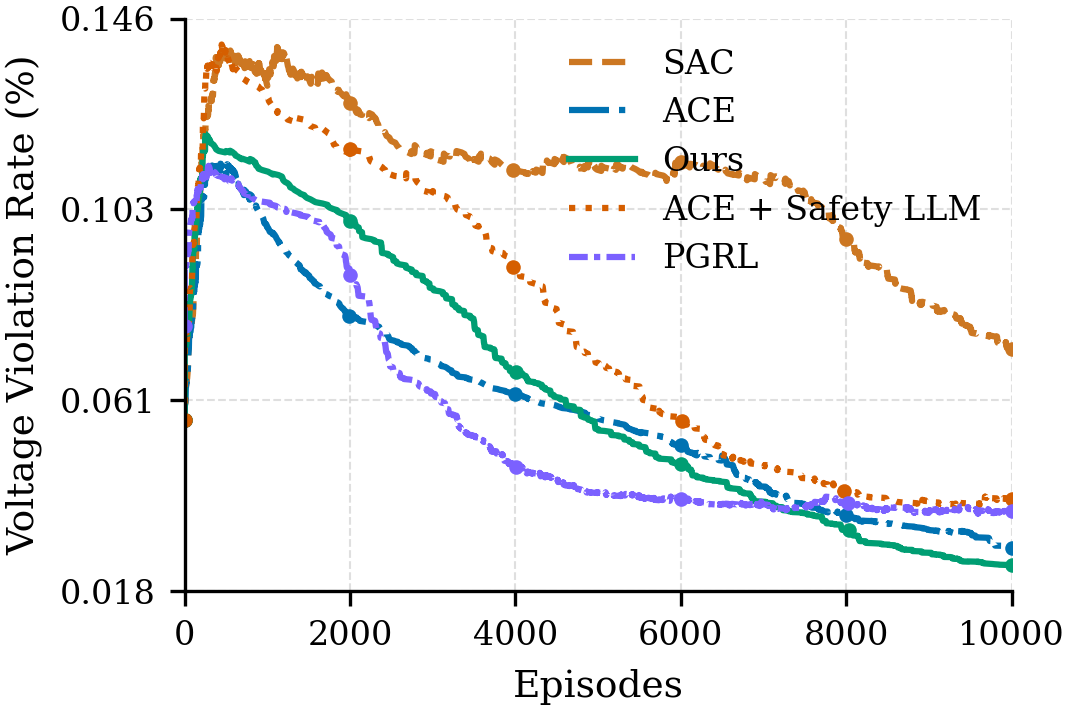}
    \caption{Voltage violation rate of the IEEE 36-bus system.}
    \label{fig:36bus-voltage}
\end{figure}
\begin{figure}[t]
    \centering
    \includegraphics[width=0.6\linewidth]{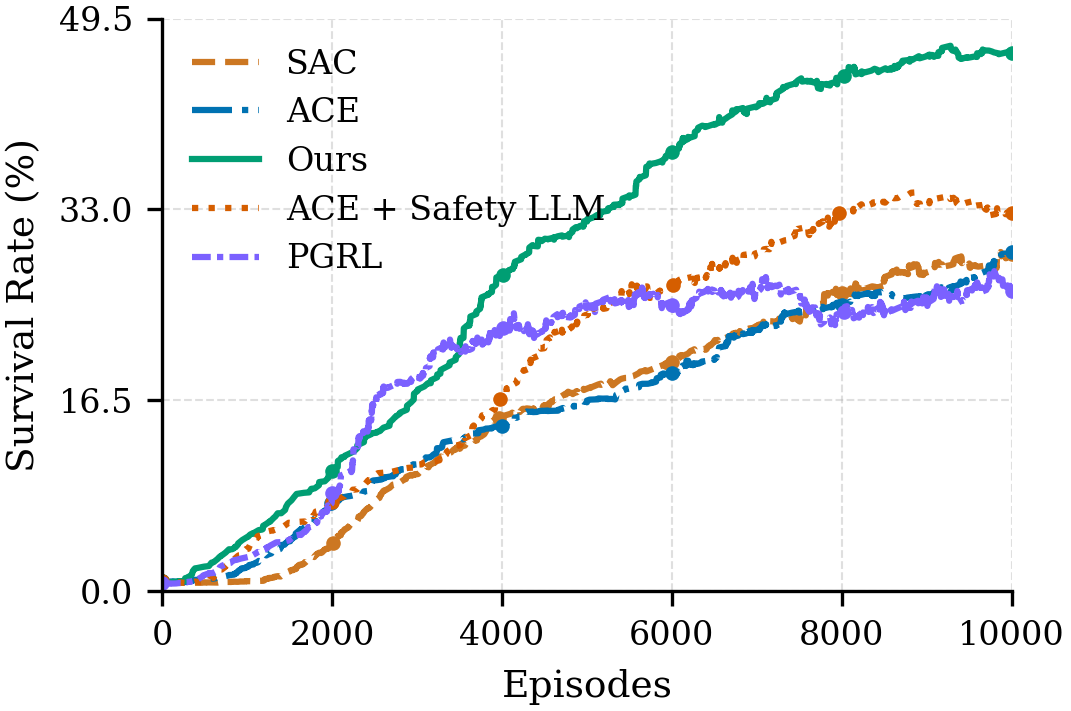}
    \caption{Survival Rate on 36-bus}
    \label{fig:36bus-step}
\end{figure}

\subsubsection{Training Dynamics and Convergence Trends}
\label{sec:trends36}

Fig.~\ref{fig:36bus-step}, ~\ref{fig:36bus-overload} and~\ref{fig:36bus-voltage} illustrate the training evolution.
Our method improves survival duration rapidly while maintaining stable growth, in contrast to the oscillatory behavior of SAC and ACE.
Similarly, the overload and voltage violation rates decrease earlier and remain lower for our full model.
\textcolor{blue}{PGRL exhibits fast initial reduction in overload and safety cost, but converges to a lower performance ceiling, reflecting the limited adaptability of purely heuristic strategies.}
Although ACE with Safety-SAC stabilizes faster than ACE, it still lags behind the full model, confirming that the combined use of Safety-SAC and Safety LLM ensures both stable convergence and safe exploration.

{\color{red}The differing behavior of Safety-SAC across the two systems may stem from structural differences. In the 36-bus system, the action space is relatively small, and constraint violations may lead to sharp reward fluctuations, making the safety critic more sensitive and occasionally over-penalizing certain transitions. In contrast, the 118-bus system has a substantially larger combinatorial action space, where the safety critic provides more effective guidance by filtering high-risk regions during exploration. This scale effect partially explains why Safety-SAC shows clearer advantages in the larger system.}

\subsection{Results on IEEE 118-bus System}

Tables~\ref{tab:wcci_reward} and~\ref{tab:wcci_safety} present the evaluation results on the IEEE 118-bus system. Owing to the significantly larger state--action space, this benchmark highlights the scalability of our framework.
\begin{table}[htbp]
  \centering
  \caption{The RL episode reward and survival step in IEEE 118-bus case.}
  \begin{tabular}{lcc}
    \toprule
    Method & Reward $\times10^{5}$ $\uparrow$ & Survival Step $\uparrow$ \\
    \midrule
    PGRL&0.8197&69.1\\
    SAC & 1.1018 & 91.9 \\
    ACE & 2.2428 & 189.9 \\
    ACE with Safety-SAC & 3.1684 & 271.1 \\
    Ours & \textbf{4.2898} & \textbf{357.4} \\
    \bottomrule
  \end{tabular}
  \label{tab:wcci_reward}
\end{table}

\begin{table}[htbp]
  \centering
  \caption{Safety metrics in the IEEE 118-bus case.}
  \begin{tabular}{lccc}
    \toprule
    Method & Overload (\%) $\downarrow$ & Voltage (\%) $\downarrow$ & SafeCost $\downarrow$ \\
    \midrule
    PGRL&0.942&4.982&1.024\\
    SAC & 1.033 & 4.341 & 1.364 \\
    ACE & 0.845 & 6.788 & 0.964 \\
    ACE with Safety-SAC & 0.486 & 3.108 & 0.748 \\
    Ours & \textbf{0.418} & \textbf{1.731} & \textbf{0.549} \\
    \bottomrule
  \end{tabular}
  \label{tab:wcci_safety}
\end{table}

The proposed method achieves the highest reward and longest survival duration, outperforming \textcolor{blue}{PGRL} SAC, ACE, and ACE with Safety-SAC by a wide margin. Although Safety-SAC enhances the stability of ACE, the Safety LLM substantially amplifies this improvement, underscoring its importance in large-scale decision spaces where unsafe exploration is more likely. \textcolor{blue}{While PGRL provides moderate safety improvements over SAC, its performance degrades noticeably in terms of reward and survival, indicating that heuristic physics-guided strategies alone do not scale well to large and highly combinatorial grids.}

Our method achieves the lowest overload and voltage violation rates, leading to the minimum SafetyCost. ACE with Safety-SAC provides only partial violation reduction and shows degradation in later episodes. 
\textcolor{blue}{PGRL reduces overload events in early training but fails to consistently control voltage violations, reflecting the limitations of fixed physics-guided heuristics under increasing system complexity.} In contrast, the full model consistently maintains low-risk operational states, indicating that LLM-guided reasoning is particularly beneficial for managing complex grids with high renewable fluctuations.

\subsubsection{Training Dynamics and Convergence Trends}
\label{sec:trends118}
\begin{figure}[t]
    \centering
    \includegraphics[width=0.6\linewidth]{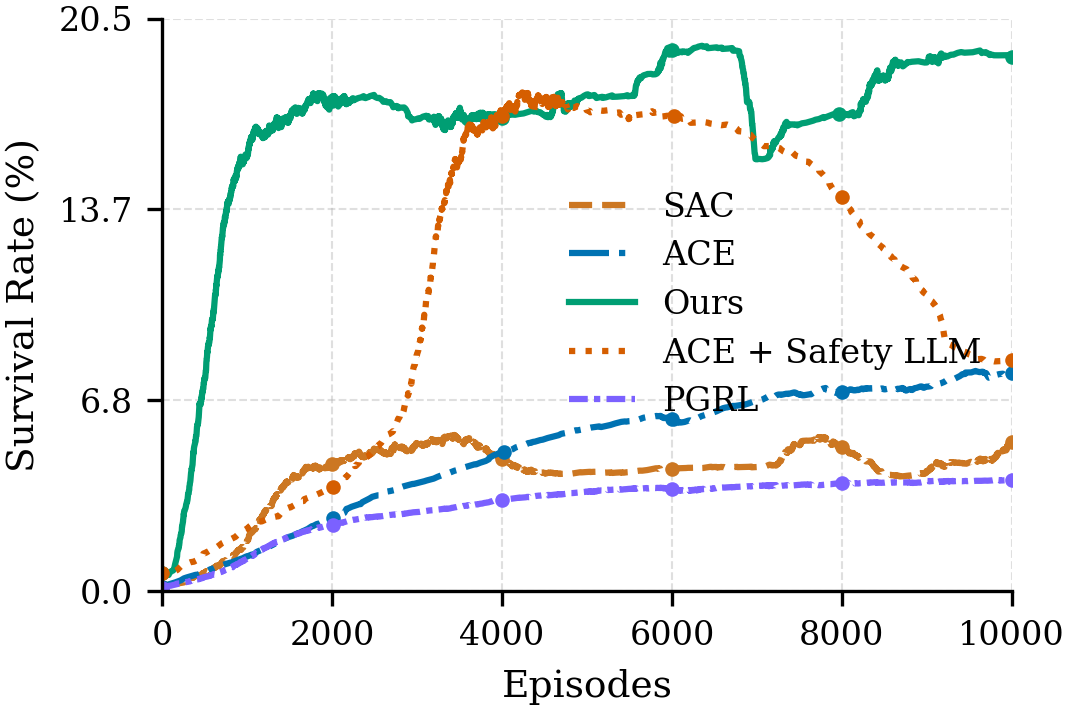}
    \caption{118-bus Survival Step}
    \label{fig:118bus-step}
\end{figure}

\begin{figure}[t]
    \centering
    \includegraphics[width=0.6\linewidth]{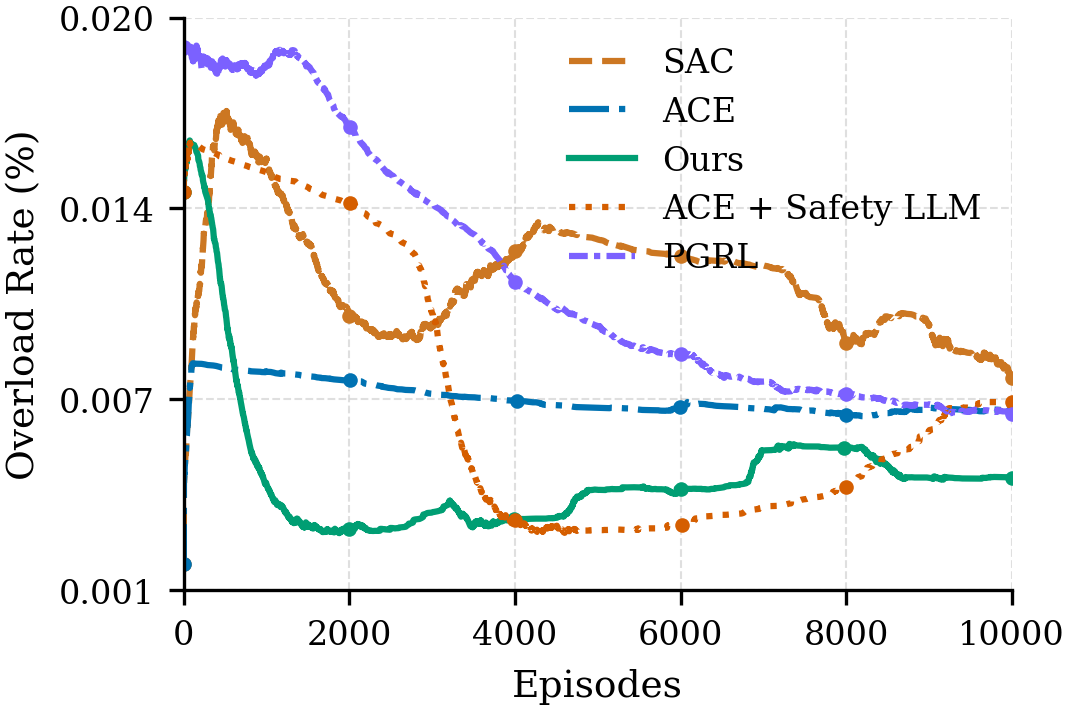}
    \caption{Overload rate of the IEEE 118-bus system.}
    \label{fig:118bus-overload}
\end{figure}

\begin{figure}[t]
    \centering
    \includegraphics[width=0.6\linewidth]{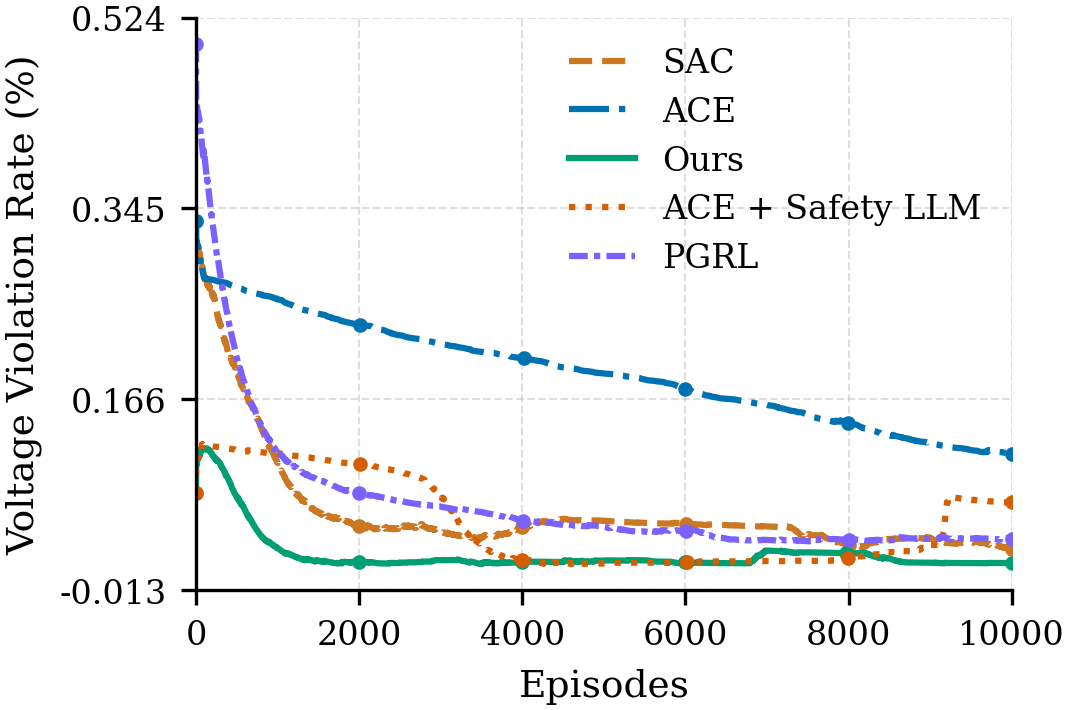}
    \caption{Voltage violation rate of the IEEE 118-bus system.}
    \label{fig:118bus-voltage}
\end{figure}

Fig.~\ref{fig:118bus-step}, ~\ref{fig:118bus-overload} and ~\ref{fig:118bus-voltage} show that our method exhibits the most stable and rapid convergence. Survival duration improves quickly, and both overload and voltage violations remain near zero for most of the training horizon. SAC and ACE converge slowly and unstably, whereas ACE with Safety-SAC shows intermediate behavior. \textcolor{blue}{PGRL converges quickly in early stages but plateaus prematurely, further confirming that physics-guided heuristics alone are insufficient for scalable and long-horizon safe control.} These trends confirm that integrating Safety-SAC with the Safety LLM yields scalable, safe, and robust learning behavior.

\subsection{Ablation Study}

To better understand the influence of the key hyperparameters in the proposed framework, we perform ablation experiments on the reward threshold $r_{\mathrm{thr}}$ and the refinement frequency $f$. The corresponding results are shown in Fig.~\ref{fig:ab_thr} and Fig.~\ref{fig:ab_f}. These two hyperparameters respectively determine which transitions are flagged as suboptimal and how often the LLM-based refinement is invoked, both of which directly influence the strength and stability of knowledge-based guidance.

\subsubsection{Effect of Reward Threshold $r_{\mathrm{thr}}$}

\begin{figure}[t]
    \centering
    \includegraphics[width=0.5\linewidth]{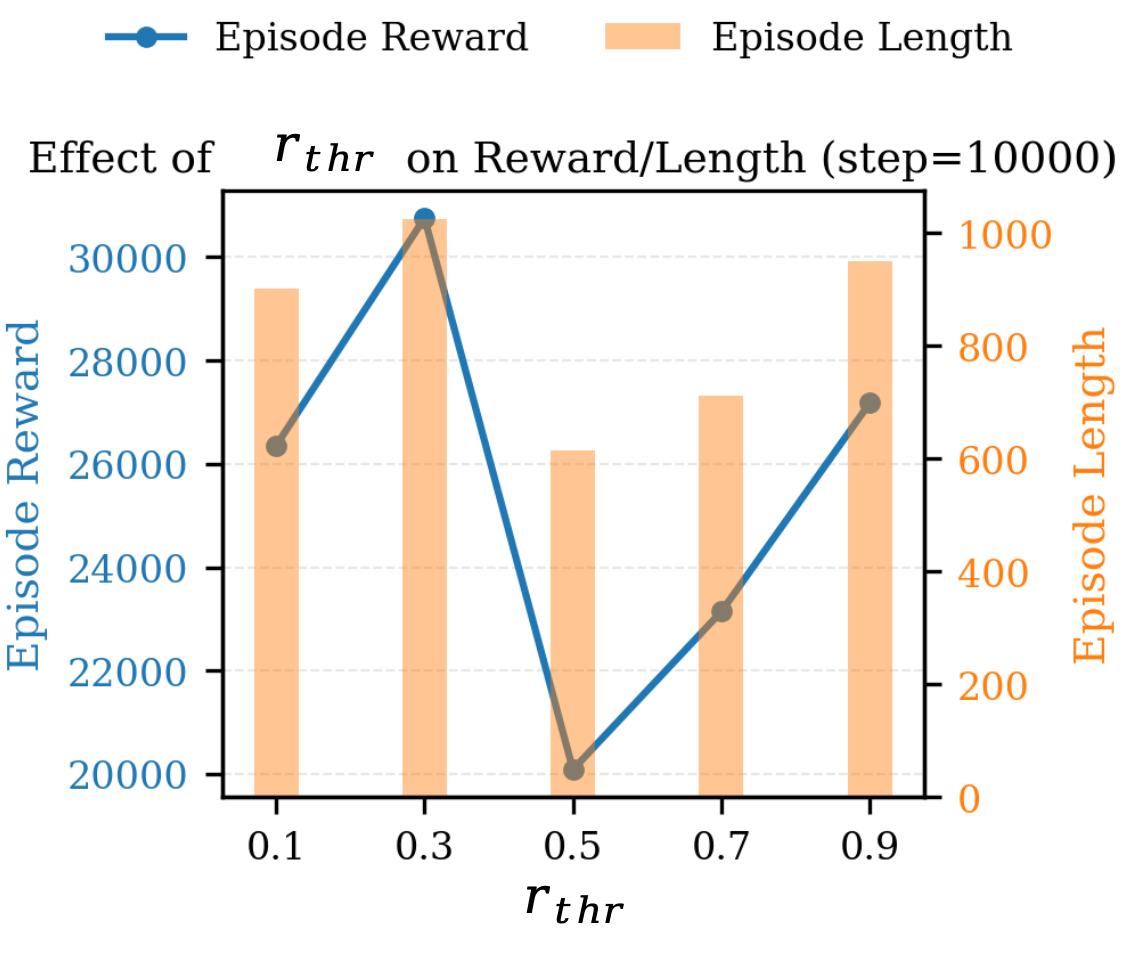}
    \caption{Ablation result for the reward threshold $r_{\mathrm{thr}}$.}
    \label{fig:ab_thr}
\end{figure}

Fig.~\ref{fig:ab_thr} illustrates how different values of $r_{\mathrm{thr}}$ affect the episode reward and episode length. The best overall performance is achieved when $r_{\mathrm{thr}} = 0.3$, where the agent attains both the highest cumulative reward and the longest survival duration. When the threshold is too small (e.g., $0.1$), only a limited number of transitions fall below the threshold, causing the LLM to intervene infrequently. This weakens the corrective effect and provides insufficient knowledge-based guidance during exploration. In contrast, larger thresholds (e.g., $0.5$ or above) classify a significant portion of transitions as suboptimal, causing the LLM to intervene excessively. Such frequent refinements interrupt the underlying RL learning dynamics and lead to instability, eventually deteriorating the achieved reward and survival length. These results suggest that an intermediate threshold effectively balances the depth of refinement and the stability of policy learning.

\subsubsection{Effect of Refinement Frequency $f$}

\begin{figure}[t]
    \centering
    \includegraphics[width=0.6\linewidth]{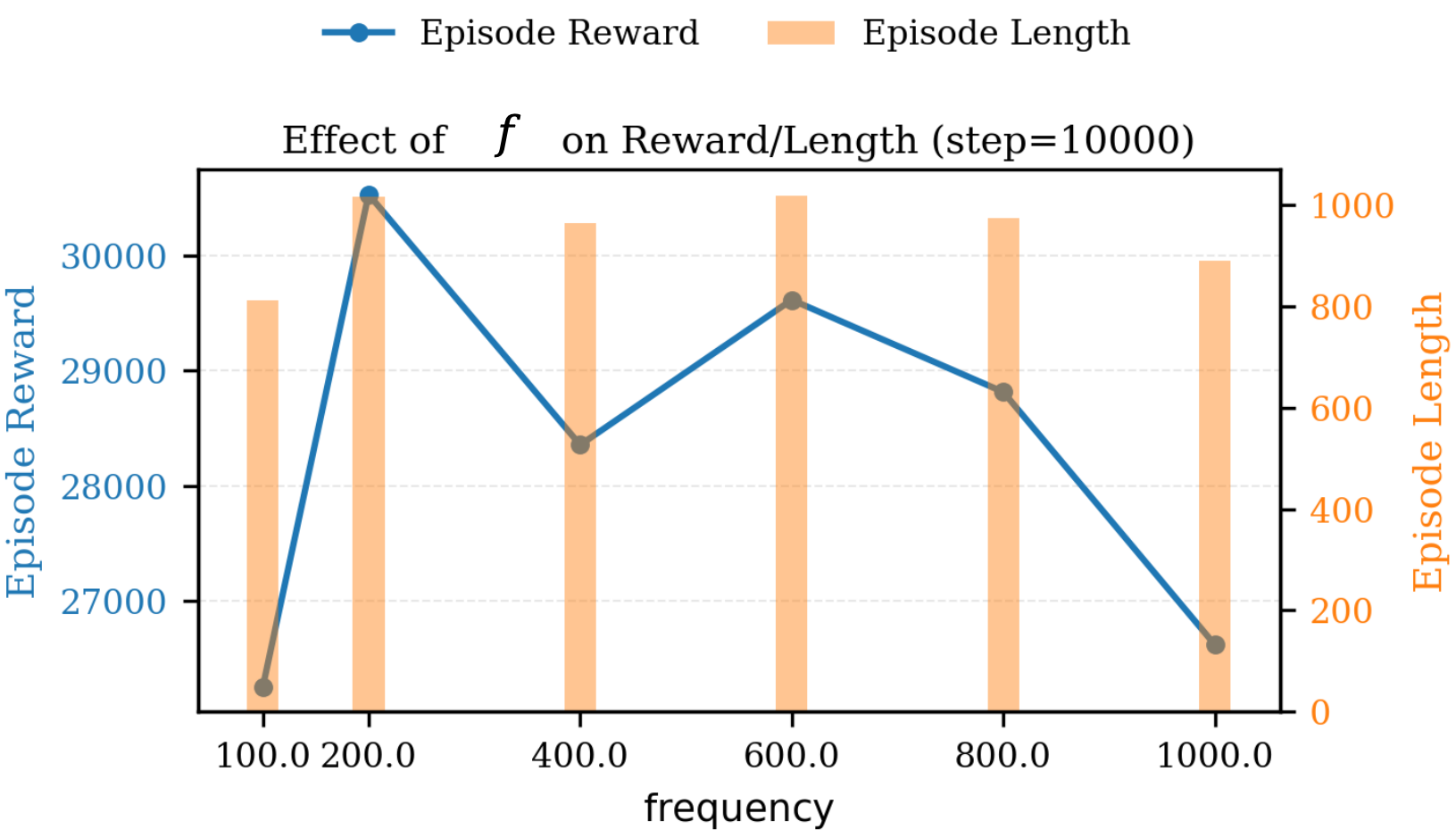}
    \caption{Ablation result for the refinement frequency $f$.}
    \label{fig:ab_f}
\end{figure}

The impact of the refinement interval $f$ is shown in Fig.~\ref{fig:ab_f}. When $f$ is small, the LLM is invoked very frequently, resulting in overly aggressive modifications to the agent’s actions. This leads to oscillatory behavior and prevents the policy from settling into a stable operating regime, thereby producing relatively low rewards and short episode lengths. As $f$ increases, the refinement becomes less intrusive, allowing the RL backbone to accumulate more consistent experience between consecutive corrections. The best performance is reached at $f = 200$, indicating that this frequency provides the most effective balance between refinement strength and RL autonomy. However, further increasing $f$ to $800$–$1000$ makes the refinements too sparse. With insufficient LLM feedback, the agent struggles to correct poor behaviors in time, which reduces both reward and survival duration. These results highlight the importance of selecting an appropriate refinement interval that maintains both stability and responsiveness during training.

{\color{red}The ablation results reveal a stability–responsiveness trade-off. A low $r\_{thr}$ reduces intervention frequency but may allow suboptimal transitions to persist. A high $r\_{thr}$ increases correction intensity but may introduce training oscillation. Similarly, too frequent LLM invocation (small $f$) may destabilize learning, while infrequent updates delay corrective feedback.}

\begin{figure*}[t]
    \centering
    \includegraphics[width=\textwidth]{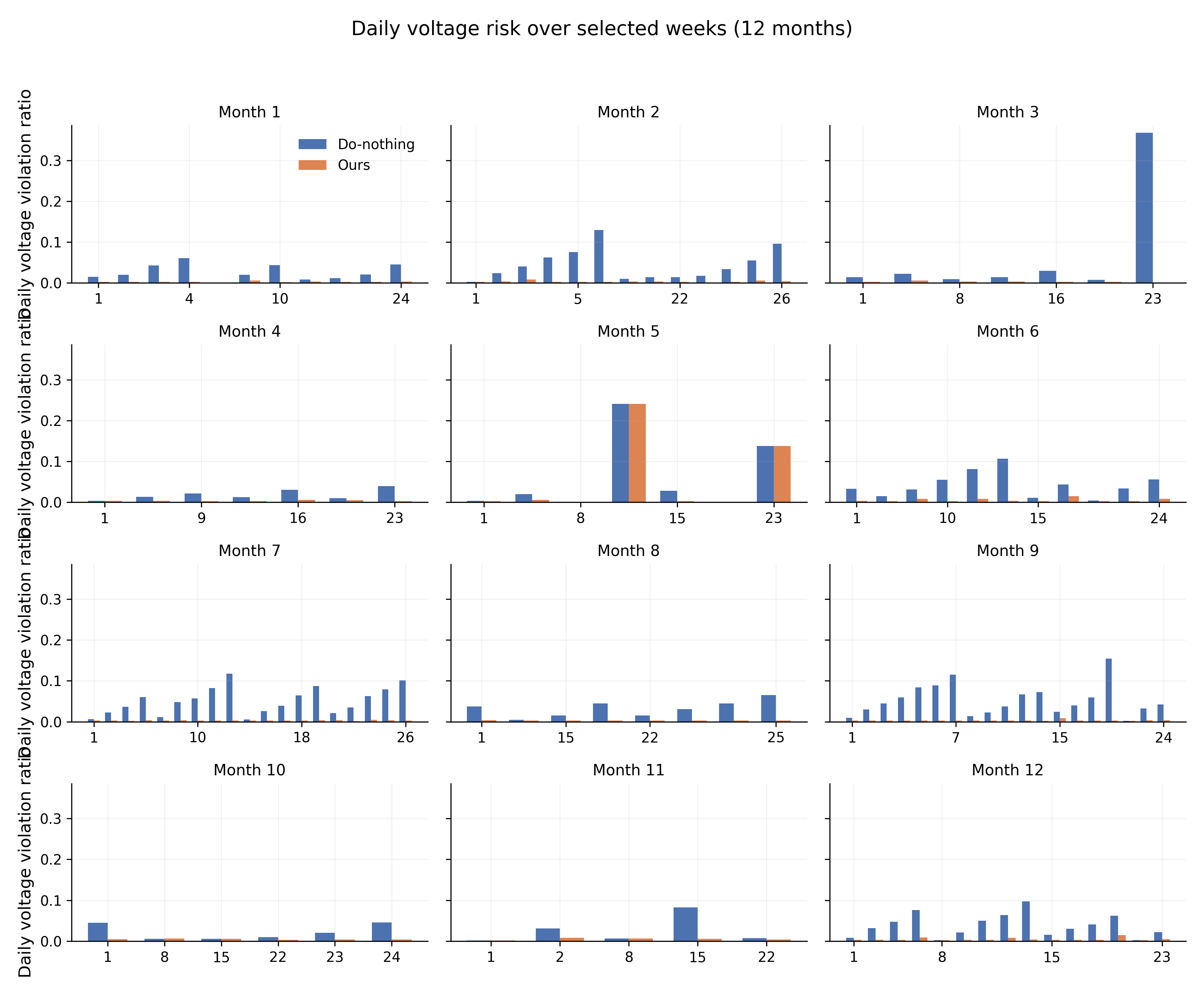}
    \caption{
    Daily voltage violation ratio across twelve months.
    Each month contains 28 representative days derived from the top-performing weeks under the RL strategy.  }
    \label{fig:daily_voltage}
\end{figure*}

\begin{figure*}[t]
    \centering
    \includegraphics[width=\textwidth]{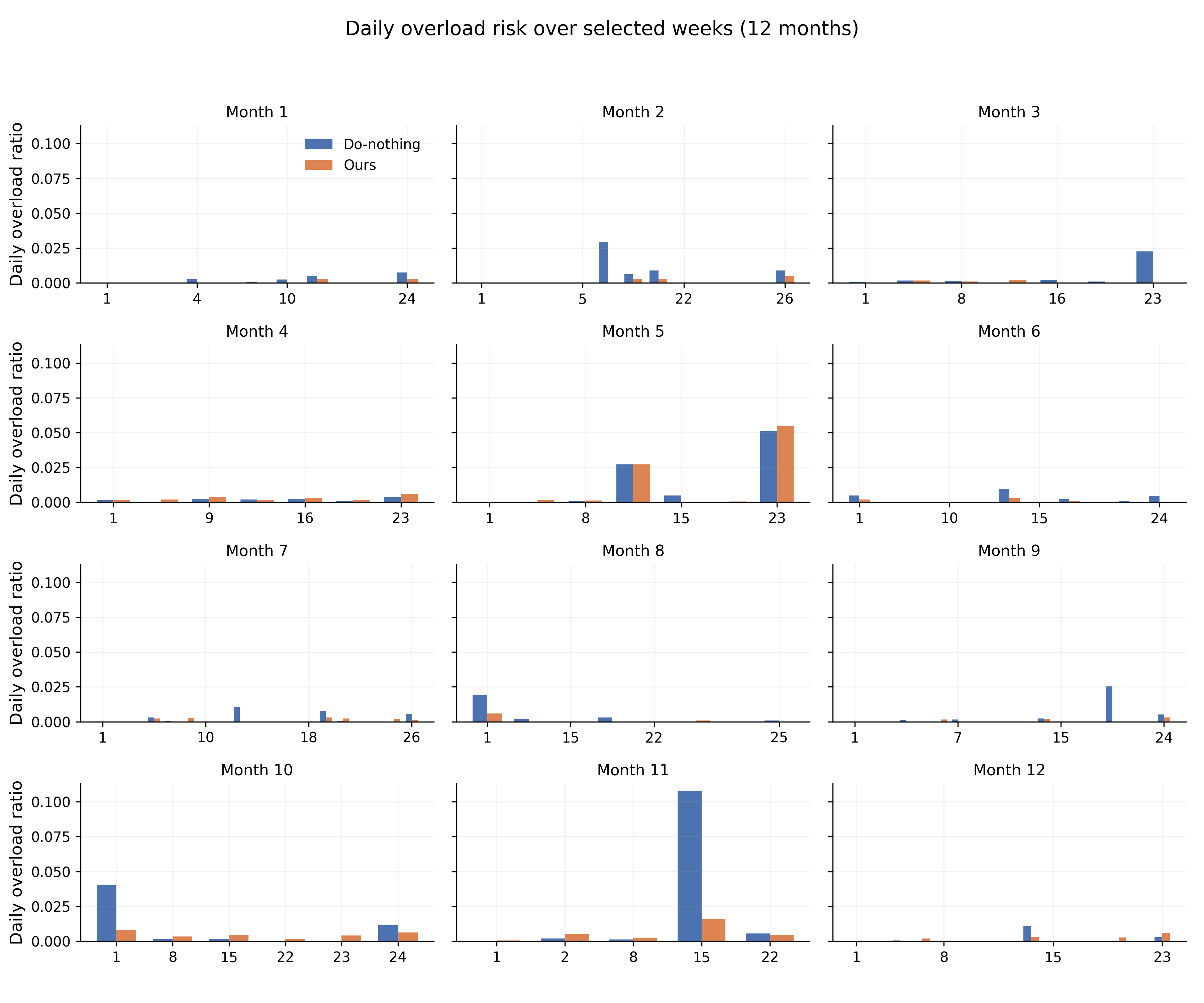}
    \caption{
    Daily overload risk distribution for all twelve months.
    }
    \label{fig:daily_overload}
\end{figure*}

\begin{figure*}[t]
    \centering
    \includegraphics[width=\textwidth]{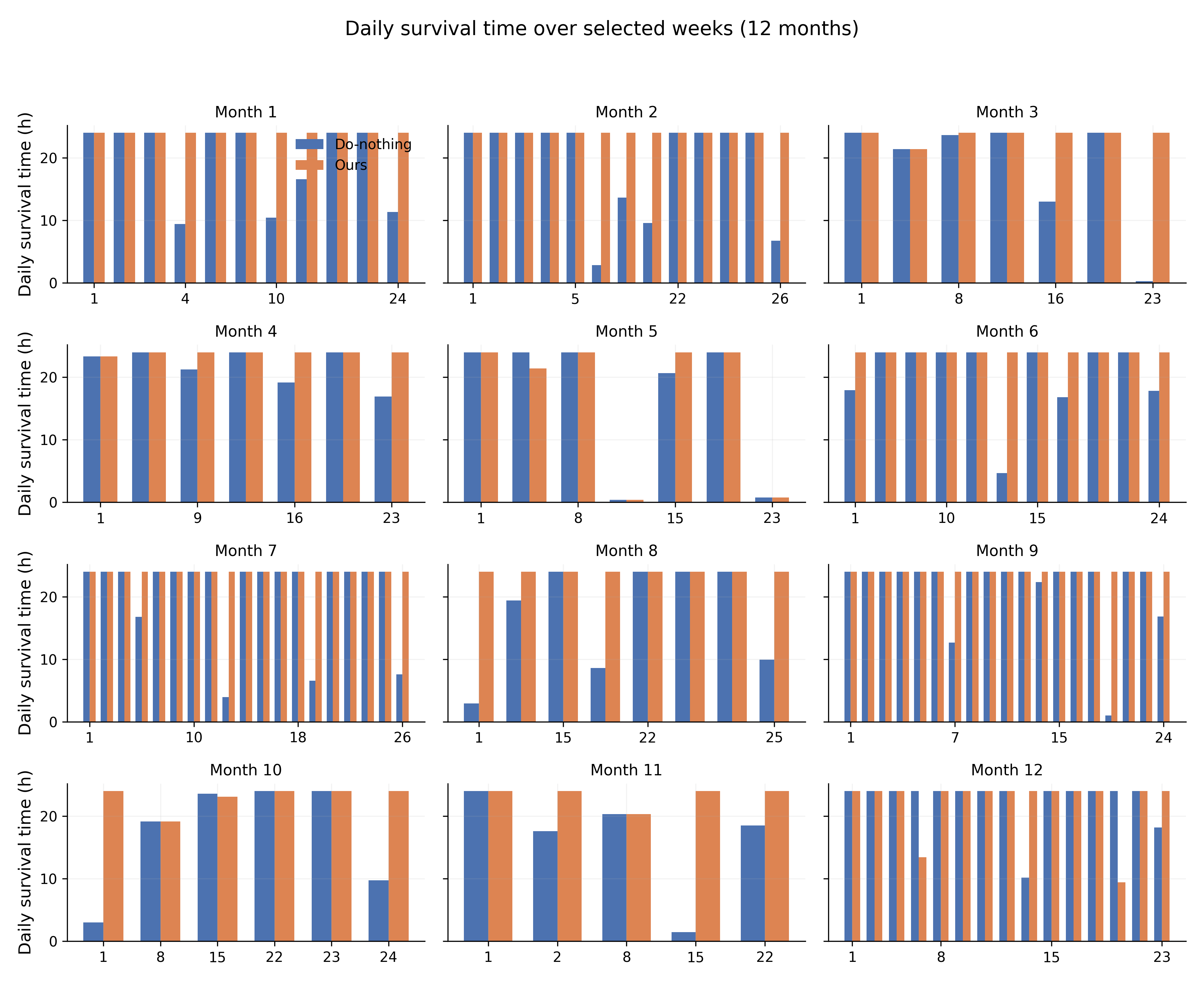}
    \caption{
    Daily survival time across twelve months.  
    }
    \label{fig:daily_survival}
\end{figure*}

\subsection{Daily Performance Across 12 Months}
Figs.~\ref{fig:daily_voltage}--\ref{fig:daily_survival} present the daily operational performance over twelve months on IEEE 36-bus system. These multi-panel figures allow a fine-grained comparison between uncontrolled system operation and the method introduced in this work.

\subsubsection{Voltage Violation Risk}
Across all months, operation without control intervention frequently leads to substantial voltage violations.
Several months (e.g., Month~3, Month~5, Month~11) show pronounced daily spikes, with violation ratios exceeding $0.3$.
In contrast, the proposed method maintains voltage violation ratios near zero for the overwhelming majority of days, even under stressed conditions.
This highlights the method’s ability to take timely corrective actions to keep voltage magnitudes within acceptable limits.

\subsubsection{Overload Risk}
A similar pattern appears in the overload risk.
Under no intervention, the system exhibits recurring overload events, especially during periods of heavy load (Month~5, Month~9, Month~11).
Conversely, the proposed method effectively suppresses overload occurrences across all months.
Even on the most challenging days, the overload ratios achieved by the proposed strategy remain substantially lower than those under uncontrolled operation.
This demonstrates the method’s capability to redistribute power flows via preventive adjustments and targeted corrective actions.

\subsubsection{Daily Survival Time}
Daily survival time further highlights the differences in system resilience.
Under uncontrolled operation, the grid frequently fails to sustain a full day of operation, with collapses sometimes occurring only a few hours into the simulation.
The method proposed in this work, however, consistently maintains system stability for nearly the entire 24-hour horizon across all months, with only limited exceptions.
These results indicate that the proposed strategy effectively prevents cascading failures and mitigates destabilizing contingencies before they propagate.

\subsection{Monthly Average of Weekly Metrics}

\begin{figure}[t]
    \centering
    \includegraphics[width=\columnwidth]{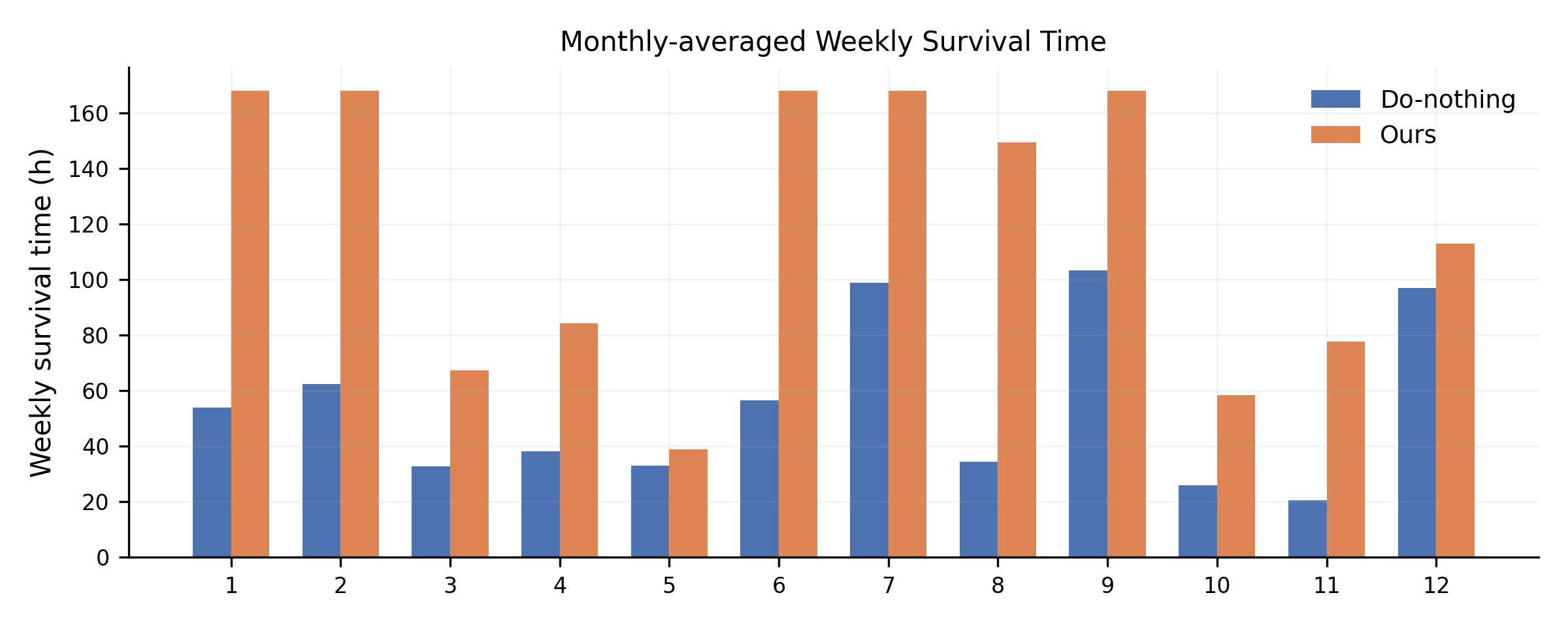}
    \caption{
    Monthly-averaged weekly survival time.  
    }
    \label{fig:yearly_survival}
\end{figure}

\begin{figure}[t]
    \centering
    \includegraphics[width=\columnwidth]{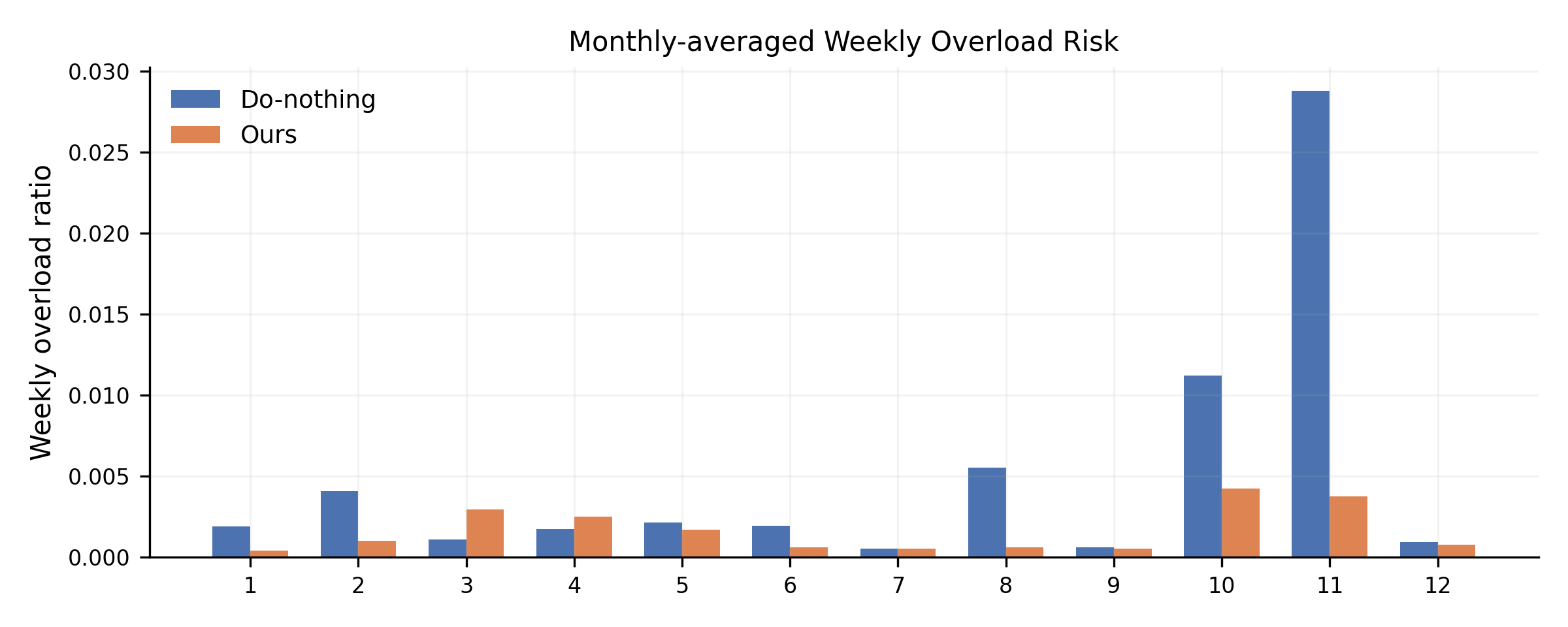}
    \caption{
    Monthly-averaged weekly overload risk for the baseline and the proposed method.
    }
    \label{fig:yearly_overload}
\end{figure}

\begin{figure}[t]
    \centering
    \includegraphics[width=\columnwidth]{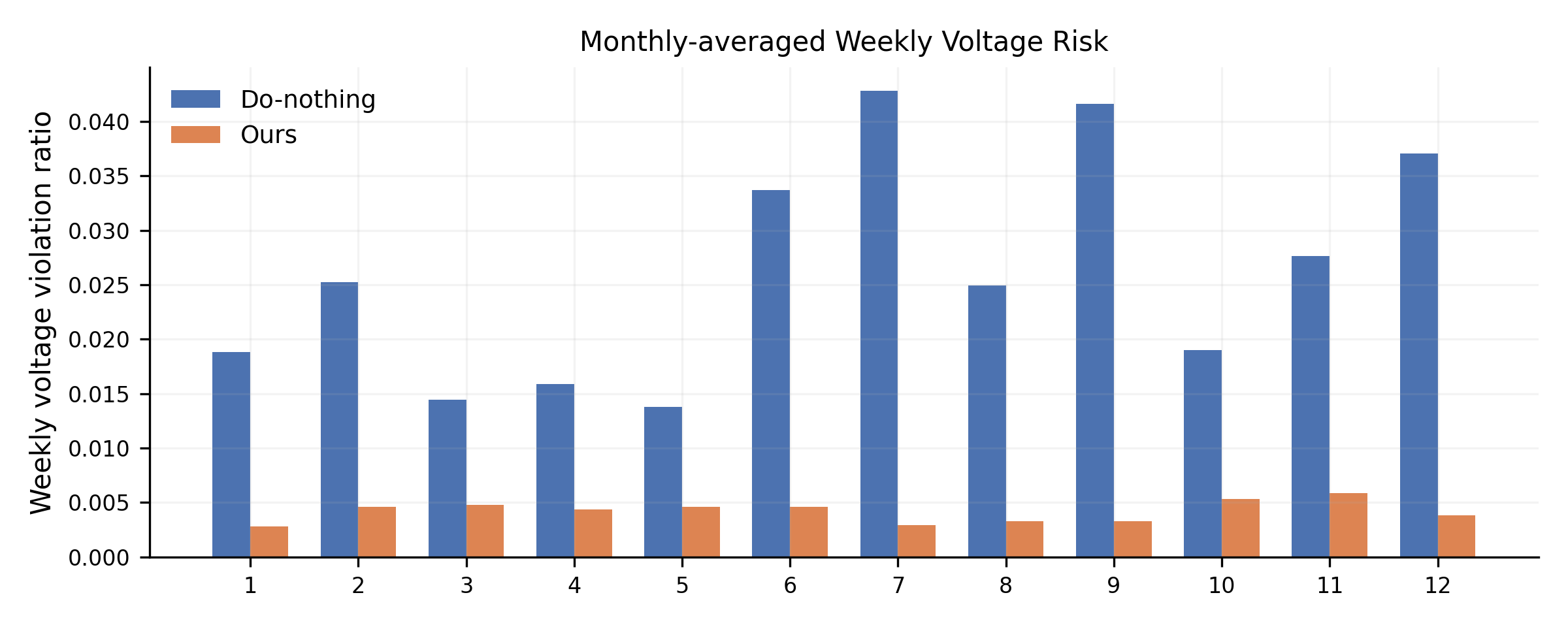}
    \caption{
    Monthly-averaged weekly voltage violation ratio.  
    }
    \label{fig:yearly_voltage}
\end{figure}

To evaluate long-term operational robustness, the weekly performance metrics are averaged over the representative weeks from each month on IEEE 36-bus system.
The results shown in Figs.~\ref{fig:yearly_survival}--\ref{fig:yearly_voltage} demonstrate the clear advantages of the proposed control strategy.

\subsubsection{Weekly Survival Time}

Under the condition of electric grid operation without any control intervention, the system typically maintains only $30$–$100$ hours of stable operation per week (out of a maximum of $168$ hours).
In contrast, the proposed method consistently sustains $160$–$170$ hours per week throughout the entire year.
The improvement is especially significant in the more challenging months (e.g., Month~1, Month~2, Month~6, Month~7), where uncontrolled operation collapses early while the proposed method keeps the system stable for nearly the full duration.

\subsubsection{Weekly Overload Risk}

The uncontrolled operation exhibits monthly overload risks between $0.005$ and $0.03$, with notable peaks in Months~7 and~11.
By comparison, the proposed method effectively suppresses overload risk to near-zero across all months.
This demonstrates the method’s strong capability in managing thermal constraints and avoiding unsafe line loading.

\subsubsection{Weekly Voltage Violation Risk}

Monthly voltage violation ratios under no-intervention operation remain relatively high, ranging from $0.015$ to $0.043$.
With the proposed method, voltage violations are consistently reduced below $0.005$, representing approximately a $70\%$–$90\%$ reduction across all months.
These results confirm that the proposed control approach can maintain stable voltage profiles even under highly variable operating conditions.

\begin{figure}[t]
    \centering
    \includegraphics[width=0.7\linewidth]{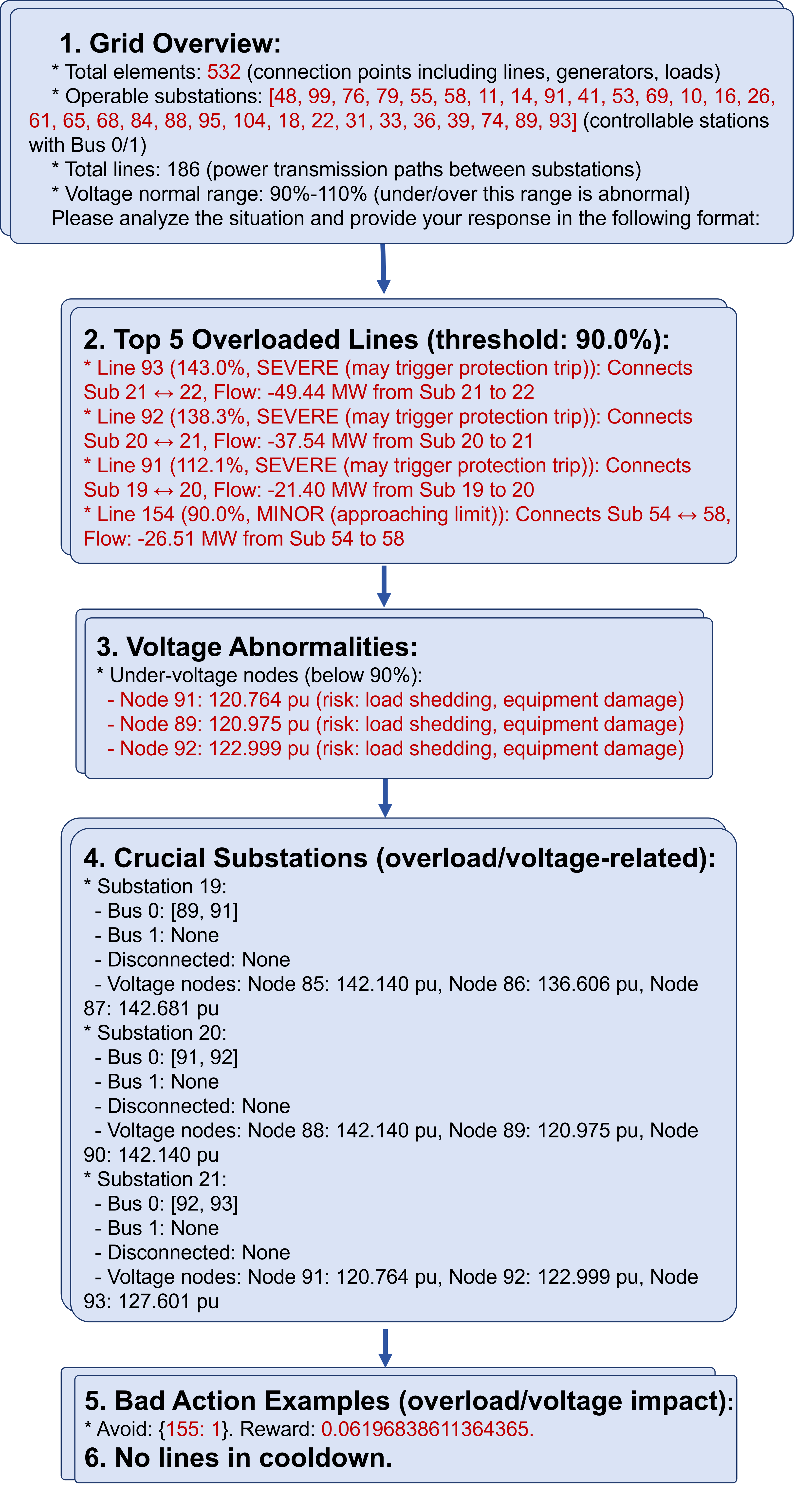}
    \caption{\textcolor{blue}{Visualization of the topology action prompt proposed by the RL policy for a representative IEEE 118-bus state (input).}}
    \label{fig:llm_refine_input}
\end{figure}

\begin{figure}[t]
    \centering
    \includegraphics[width=0.7\linewidth]{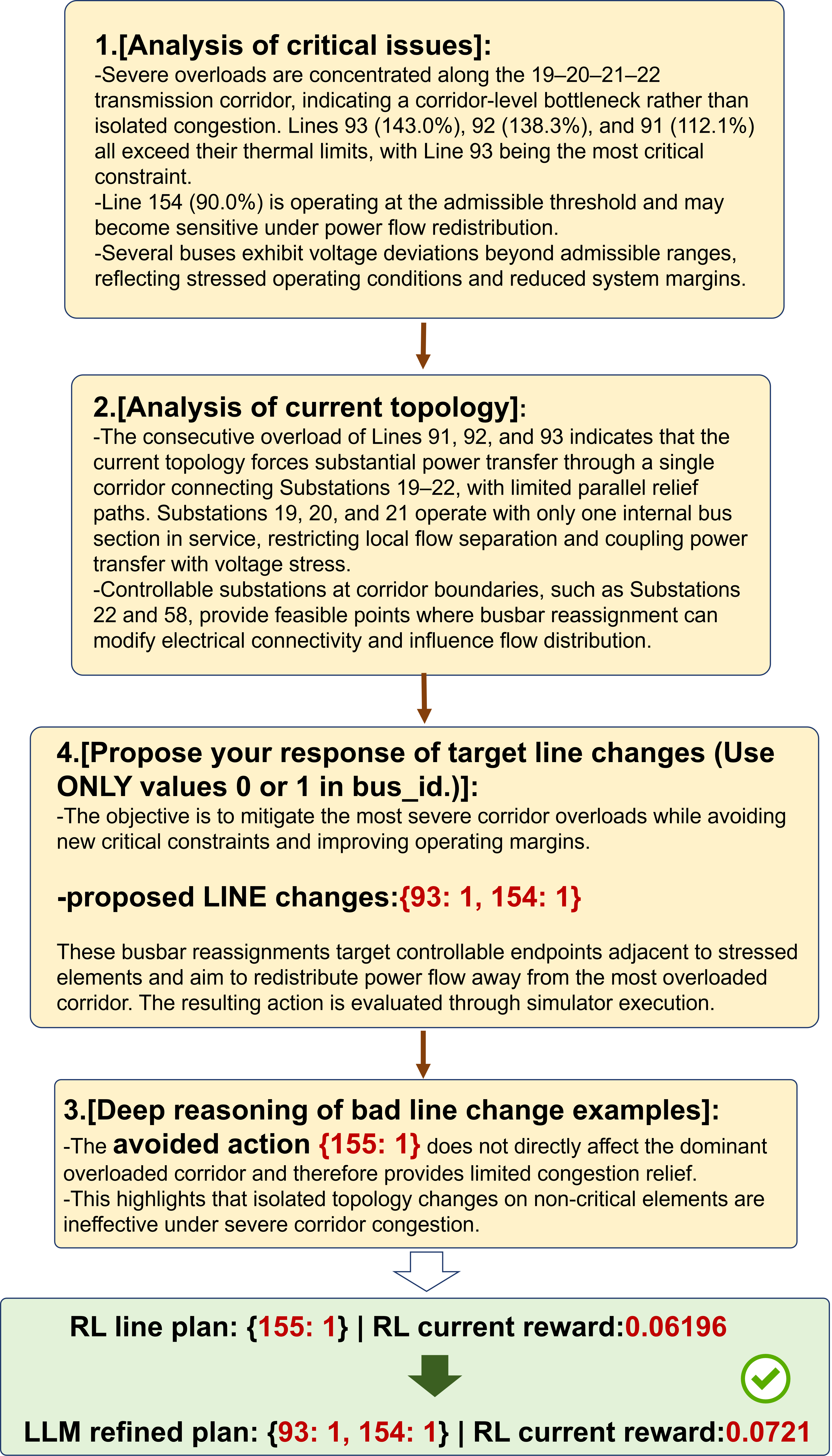}
    \caption{\textcolor{blue}{Visualization of the topology action refined by the LLM for the same IEEE 118-bus state (output). The refined action yields a higher immediate return in the simulator and is stored as an additional transition in the replay buffer.}}
    \label{fig:llm_refine_output}
\end{figure}

{\color{blue}
\subsection{LLM-based transition refinement and replay-buffer effects}

This subsection provides an experimental illustration of how LLM-based transition refinement interacts with the replay buffer and, consequently, the learning dynamics of Safety-SAC. Figs.~\ref{fig:llm_refine_input} and~\ref{fig:llm_refine_output} show a representative refinement instance in the IEEE 118-bus environment. Starting from the identical system state \(s\), the baseline policy proposes an admissible topology control action \(a_{\pi}\) (Fig.~\ref{fig:llm_refine_input}), which results in a low immediate return due to insufficient mitigation of security-related penalties embedded in the environment reward. In contrast, the LLM proposes an alternative action \(a_{\mathrm{LLM}}\) (Fig.~\ref{fig:llm_refine_output}) that is successfully parsed into a valid Grid2Op command and, when evaluated by the simulator, produces a higher immediate return.

Importantly, refinement is applied sparsely under predefined triggering criteria: the LLM is invoked periodically (every \(f\) steps) and only for transitions whose immediate reward falls below the threshold \(r_{\mathrm{thr}}\). For each triggered case, the original transition \((s,\ a_{\pi},\ r,\ s')\) is preserved, and an additional refined transition \((s,\ a_{\mathrm{LLM}},\ r^{*},\ s^{*})\) is appended to the replay buffer. Therefore, the method does not overwrite past experience or alter the simulator dynamics; instead, it augments the buffer with a limited set of corrective, physically feasible transitions that are reachable from the same states.

From a learning perspective, this augmentation locally enriches the conditional action coverage \(p(a \mid s)\) around challenging operating points, where the base policy is more likely to generate suboptimal or unsafe switching decisions. By repeatedly presenting the critic and actor updates with alternative actions that achieve improved security-aware returns under comparable conditions, the refinement mechanism improves sample efficiency and stabilizes policy updates. The resulting reduction in oscillatory learning behavior and the improved reliability/safety metrics observed in the learning curves are consistent with the targeted experience-augmentation principle described in Section~3.5.}

\section{Conclusions and future work}

This study proposes an LLM-guided safe reinforcement learning framework for power system topology reconfiguration, aiming to improve operational reliability under uncertain and dynamic conditions. By introducing differentiable safety-cost signals and a Safety-SAC structure, the agent can learn stable policies while respecting voltage and thermal limits. Moreover, the Safety-LLM provides adaptive action refinement, allowing the learning process to incorporate system-level reasoning rather than relying solely on numerical critics. Evaluations on the IEEE 36-bus and 118-bus systems confirm that the proposed framework improves both performance and safety compared with SAC, ACE, and their safety-enhanced variants.

Based on comparative experiments, some key findings are obtained as follows:
(1) Safety improvement: The proposed method significantly reduces overload and voltage violation rates, indicating more effective constraint handling during exploration and operation.
(2) Operational performance: The framework achieves higher cumulative rewards and longer survival times, showing that improved safety does not come at the cost of operational efficiency.
(3) Scalability: Performance gains remain consistent when moving from the 36-bus to the 118-bus system, demonstrating that LLM-guided safe RL can handle large action spaces and complex grid dynamics.



{\color{red}Future work will extend the framework to broader security constraints, such as transient stability and frequency limits, and evaluate its performance under multi-energy and forecasting-integrated settings. While a representative reasoning-capable LLM is adopted in this study, systematic evaluation across different architectures and model scales remains future work. In practical deployment, the framework can function as a decision-support module within digital-twin or operator-in-the-loop environments, where refined actions are validated through existing simulators, preserving compatibility with current energy management systems.}

\bibliographystyle{elsarticle-num}  
\bibliography{ref}

\end{document}